\documentclass[12pt]{article}
\usepackage{amsmath,amssymb,epsfig,ulem}
\usepackage[bookmarks=false]{hyperref}
\usepackage{color}
\usepackage{multicol}

\allowdisplaybreaks

\makeatletter

\newenvironment{figurehere}
{\def\@captype{figure}}
{}
\makeatother

\renewcommand{\arraystretch}{1.2}

\newcommand{\beq}{\begin{equation}}
\newcommand{\eeq}{\end{equation}}
\newcommand{\bea}{\begin{eqnarray}} 
\newcommand{\eea}{\end{eqnarray}}

\newcommand{\va}{V\hspace{-1mm}A}

\begin{document}

\begin{titlepage}

\begin{flushright}
MZ-TH/12-37\\
HIM-2012-03
\end{flushright}

\vspace{15pt}
\begin{center}
  \Large\bf Top-Quark Charge Asymmetry with a Jet Handle
\end{center}

\vspace{5pt}
\begin{center}
{\sc Stefan Berge$^{a}$ and Susanne Westhoff$^{a,b}$}\\
\vspace{10pt} {\sl
$^{a}$ PRISMA Cluster of Excellence, Institut f\"ur Physik (WA THEP), \\
Johannes Gutenberg-Universit\"at,
D-55099 Mainz, Germany} \\
\vspace{10pt} {\sl
$^b$ Helmholtz-Institut Mainz,  
Johannes Gutenberg-Universit\"at \\
D-55099 Mainz, Germany} \\
\end{center}

% Pairs of top and antitop quarks are produced at the LHC to a large extent in association with a hard jet. We investigate the charge asymmetry in $t\bar t + j$ production in quantum chromodynamics and with massive color-octet vector bosons. The total charge asymmetry at the LHC is suppressed by the large charge-symmetric background from gluon-gluon fusion. We show to what extent the asymmetry can be enhanced by suitable phase space cuts. In particular, we elaborate on the jet kinematics in the $t\bar t + j$ final state. Massive color octets can generate sizeable effects in $t\bar t + j$ production, and jet kinematics can be used to probe both vector and axial-vector couplings to quarks.

\vspace{10pt}
\begin{abstract}
\vspace{2pt} 
\noindent
Pairs of top and antitop quarks are produced at the LHC to a large extent in association with a hard jet. We investigate the charge asymmetry in $t\bar t + j$ production in quantum chromodynamics (QCD) and with additional massive color-octet vector bosons. The total charge asymmetry at the LHC is suppressed by the large charge-symmetric background from gluon-gluon fusion. We show to what extent the asymmetry can be enhanced by suitable phase space cuts and, in particular, elaborate on the kinematics of the hard jet in the $t\bar t + j$ final state. We demonstrate that in QCD, the asymmetry amounts to $−1.5\,\%$ for central jets without an excessive reduction of the cross section. By applying additional kinematical cuts, the asymmetry can be enhanced to $−4\,\%$, but at the cost of a strong reduction of the cross section. Massive color-octet states can generate sizeable effects in $t\bar t + j$ production, both on the charge asymmetry and on the cross section. The charge asymmetry probes both vector and axial-vector couplings to quarks. We show that massive color octets can generate asymmetries up to $\pm 10\,\%$ for moderate and up to $\pm 30\,\%$ for strong kinematical cuts to be used in experimental analyses at the LHC. Jet kinematics can be used to obtain further information about the nature of the couplings and thereby to discriminate between different models.

% {\color{blue}
% Pairs of top and antitop quarks are produced at the LHC to a large extent in 
% association with a hard jet. We investigate the charge asymmetry in $t\bar t + j$ production 
% in quantum chromodynamics (QCD) as well as for the case of 
% massive color-octet vector bosons. 

% The total charge asymmetry at the LHC is suppressed by the large charge-symmetric 
% background from gluon-gluon fusion. We show to what extent the asymmetry can be 
% enhanced by suitable phase space cuts and, in particular, elaborate on the jet kinematics 
% in the $t\bar t + j$ final state. 
% We demonstrate that in QCD, the asymmetry can be as large as $-1.5 \% $  for central jets
% without  extensive reduction of the cross section.  Applying additional kinematical cuts, 
% the asymmetry can even reach $-4\%$ but with the cost of a strong 
% reduction of the cross section.

% Additional massive color octet states can generate sizeable effects in $t\bar t + j$ production, 
% both for the charge asymmtery and for the cross section. The charge asymmetry 
% can be used to probe both vector and axial-vector couplings to quarks.
% In addition, the jet kinematics can be used to obtain further information
% about the coupling nature. We show that such massive gluons
% can generate asymmetries up to  $\pm 10 \%$ for moderate 
% and up to $\pm 30 \%$ for strong kinematical cuts used in the
% experimental analysis.
% }
\end{abstract}

\end{titlepage}

%\tableofcontents

\section{Introduction}
The year 2012 was predicted to be the year of the Higgs boson. Indeed, its recent discovery at the CERN Large Hadron Collider (LHC) marks an important step towards understanding electroweak symmetry breaking -- and the manifestation of the hierarchy problem. To date, it remains unclear how the electroweak scale is stabilized against large corrections from physics at higher scales. To solve this puzzle, it will be crucial to clarify whether there are new dynamics not far above the electroweak scale that trigger the Higgs mechanism. The top quark could be a messenger of such new dynamics. Due to its large mass, the top quark is suspected to take part in electroweak symmetry breaking and/or to interact strongly with new particles. Thus, probing the properties of the top quark and its couplings to the particles of the Standard Model (SM) might deliver insight into physics beyond the electroweak scale.

The experiments at the Tevatron and the LHC provide the appropriate environment for top-quark physics. With the data accumulated to date, accurate measurements of the cross section of top-quark pair production have been achieved at the center-of-mass (CM) energies of $\sqrt{s} = 1.96$, $7$, and $8\,\text{TeV}$ \cite{ATLAS:xsec,CMS:xsec}. The results are in good agreement with the equally precise SM predictions \cite{Langenfeld:2009wd,Ahrens:2011px}. Searches for new resonances in the top-antitop invariant mass spectrum both at the Tevatron and the LHC also did not reveal any hints to physics beyond the SM \cite{Aaltonen:2011ts,Abazov:2011gv,ATLAS:2012tx,CMS:ttboost}. The situation is less unambiguous when considering the top-quark charge asymmetry. The Tevatron collaborations CDF and D{\O} have measured the charge asymmetry in terms of a forward-backward asymmetry \cite{CDF:afbt,Abazov:2011rq} and found it to be larger than its SM prediction \cite{Kuhn:2011ri,Ahrens:2011uf}. Measurements of the correlated charge asymmetry at the LHC by the ATLAS and CMS collaborations do not show any deviation from the SM within the (large) experimental errors \cite{ATLAS:ac,CMS:ac}. Since the charge asymmetry at the LHC is very small, achieving a more precise measurement is a challenging endeavor. The Tevatron thus leaves us with a puzzle in top-quark physics that is difficult to resolve by measuring inclusive observables at the LHC.

The year 2012 will certainly also be the year of the top quark. At the LHC, top-antitop pairs are currently produced in abundance in proton-proton ($pp$) collisions at a CM energy of $\sqrt{s} = 8\,\text{TeV}$. Due to the enlarged phase space with respect to the Tevatron, a large fraction of top-antitop pairs are produced in association with one or more hard jets. The investigation of an additional jet in $t\bar t$ production is interesting for very different reasons. Firstly, the process $pp\rightarrow t\bar t + j$ is an important background for signatures with jets, leptons and missing transverse energy, for instance in Higgs production via vector boson fusion \cite{Rainwater:1999sd} or in searches for supersymmetric particles \cite{Mangano:2008ha}. Secondly, as the fraction of events with hard jets in inclusive $t\bar t$ production is sizeable, $t\bar t + j$ provides a good test of perturbative quantum chromodynamics at higher orders. The hadroproduction of top-quark pairs with a hard jet has been calculated in QCD to next-to-leading order (NLO) \cite{Dittmaier:2007wz,Dittmaier:2008uj,Melnikov:2010iu}. A first measurement of the $t\bar t + j$ production rate with respect to inclusive $t\bar t$ production by ATLAS \cite{ATLAS:2012ttbj} suggests that NLO corrections to $t\bar t + j$ production will soon be tested at the LHC. The effects of the top-quark decay and of parton showers in $t\bar t + j$ have been investigated in \cite{Melnikov:2011qx,Kardos:2011qa,Alioli:2011as}, yielding precise SM observables ready to face the experiment. Thirdly, the hard jet in $t\bar t + j$ production can serve to disentangle the properties of new particles that leave signatures in inclusive $t\bar t$ production. Following this line, CMS has recently performed a search for heavy $W'$ bosons in $t\bar t + j$ observables \cite{Chatrchyan:2012su}.

% The purpose of this work is to investigate the charge asymmetry in $t\bar t + j$ production. In QCD, the charge asymmetry in $t\bar t + j$ arises already at leading order (LO), contrary to inclusive $t\bar t$ production, where the asymmetry is a NLO quantity. The forward-backward asymmetry in $t\bar t + j$ production at the Tevatron is sizeable because the charge-asymmetric quark-antiquark parton channel dominates the cross section. At the LHC, however, the asymmetry is strongly suppressed by the large charge-symmetric background from gluon-gluon ($gg$) fusion. To increase the asymmetry at the LHC, deliberate cuts on the $t\bar t + j$ phase space are necessary. We will perform a detailed analysis of the jet kinematics at LO and study their impact on the charge asymmetry. Based on the results, we will derive a set of cuts that are suitable to increase the charge asymmetry at the LHC. The charge asymmetry in $t\bar t + j$ production in QCD is known to NLO and has been explored previously in terms of the integrated observables at the Tevatron \cite{Dittmaier:2008uj,Melnikov:2010iu,Alioli:2011as} and the LHC \cite{Alioli:2011as}. The extrapolation of our results to NLO should be straightforward with the NLO amplitudes at hand.

The purpose of this work is to investigate the charge asymmetry in $t\bar t + j$ production. In QCD, the charge asymmetry in $t\bar t + j$ arises already at leading order (LO), contrary to inclusive $t\bar t$ production, where the asymmetry is a NLO quantity. The forward-backward asymmetry in $t\bar t + j$ production at the Tevatron is sizeable because the charge-asymmetric quark-antiquark parton channel dominates the cross section. At the LHC, however, the asymmetry is strongly suppressed by the large charge-symmetric background from gluon-gluon ($gg$) fusion. To increase the asymmetry at the LHC, deliberate cuts on the $t\bar t + j$ phase space are necessary. We will perform a detailed analysis of the jet kinematics at LO and study their impact on the charge asymmetry. 
Based on the results,
we will derive a set of suitable  cuts  
and show that in QCD the charge asymmetry at the LHC can be increased considerably.
The charge asymmetry in $t\bar t + j$ production in QCD is known to NLO and has been explored previously in terms of the integrated observables at the Tevatron \cite{Dittmaier:2008uj,Melnikov:2010iu,Alioli:2011as} and the LHC \cite{Alioli:2011as}. The extrapolation of our results to NLO should be straightforward with the NLO amplitudes at hand.

% Beyond QCD, the charge asymmetry in $t\bar t + j$ production can be used to discriminate between different new particles that leave signatures in top-quark pair production. We will discuss in detail the contributions of massive color-octet vector bosons with vector and axial-vector couplings to quarks. These ``massive gluons'' arise in various extensions of the SM and some may provide an explanation of the enhanced forward-backward asymmetry observed in inclusive $t\bar t$ production. The dominant contributions of massive gluons to the integrated charge asymmetry in $t\bar t + j$ production at the LHC have been discussed before in \cite{Ferrario:2009ee}. Our analysis will be based on the complete set of massive gluon contributions. We will optimize the cuts on the charge asymmetry at the LHC that project on the region where the effects of massive gluons are large. Further, we will explore the prospects of using jet kinematics in $t\bar t + j$ production to determine the couplings of heavy color octets.

Beyond QCD, the charge asymmetry in $t\bar t + j$ production can be used to discriminate between different new particles that leave signatures in top-quark pair production. We will discuss in detail the contributions of massive color-octet vector bosons with vector and axial-vector couplings to quarks. These ``massive gluons'' arise in various extensions of the SM
in addition to the massless QCD gluons,
and some may provide an explanation of the enhanced forward-backward asymmetry observed in inclusive $t\bar t$ production. The dominant contributions of massive gluons to the integrated charge asymmetry in $t\bar t + j$ production at the LHC have been discussed before in \cite{Ferrario:2009ee}. Our analysis will be based on the complete set of massive gluon contributions. We will optimize the cuts on the charge asymmetry at the LHC that project on the region where the effects of massive gluons are large. Further, we will explore the prospects of using jet kinematics in $t\bar t + j$ production to determine the couplings of heavy color octets.

The paper is organized as follows. Section~\ref{sec:asymmetry-qcd} is devoted to the charge asymmetry in $t\bar t + j$ production in QCD. In Section~\ref{sec:phasespace}, we introduce a set of phase space variables that will serve to systematically study the kinematics of $t\bar t + j$ production. In Section~\ref{sec:qqbttg}, the charge asymmetry is analyzed at parton level for the process $q\bar q\rightarrow t\bar t g$. We discuss the jet kinematics in the limit of soft and collinear gluon emission and the dependence on the gluon energy. Our findings are then applied in a numerical analysis of the charge asymmetry in $t\bar t + j$ production at the Tevatron (Section~\ref{sec:afbtevatron}) and at the LHC (Section~\ref{sec:abclhc}), where we suggest strategic cuts that enlarge the asymmetry. The effects of detector cuts on the charge asymmetry are analyzed in Section~\ref{sec:jetcuts}. Section~\ref{sec:asymmetry-coloroctets} is devoted to the effects of massive color-octet bosons on the charge asymmetry in $t\bar t + j$ production. The generation of an asymmetry from vector and axial-vector contributions is discussed in Section~\ref{sec:axicontributions}. After having defined benchmark scenarios (Section~\ref{sec:benchmarks}), the numerical analysis of massive gluons in $t\bar t + j$ production is performed in Section~\ref{sec:axicuts}. In Section~\ref{sec:axicouplings}, we show how vector and axial-vector couplings can be distinguished by jet kinematics. Finally, our conclusions are drawn in Section~\ref{sec:conclusions}.

\section{Charge asymmetry in QCD}\label{sec:asymmetry-qcd}
% The top-quark charge asymmetry provides a test of charge conservation or, in theories with CP symmetry, of parity conservation in top-quark pair production. The differential charge asymmetry at a fixed point in phase space is defined as the difference between the production rate of a top-antitop pair and the rate when top and antitop are interchanged. For observables at hadron colliders, the key quantity is the partonic charge-asymmetric cross section as a function of the top-/antitop-quark scattering angle $\theta$ with respect to the direction of the incident quark $q$ in the $t\bar t$ system,\footnote{The $t\bar t$ system is the reference frame obtained by a boost of the parton CM frame along the beam axis, where the rapidities of the top and the antitop quark are of equal magnitude, but of opposite sign.}

The top-quark charge asymmetry provides a test of charge conservation or, in theories with CP symmetry, of parity conservation in top-quark pair production. The differential charge asymmetry at a fixed point in phase space is defined as the difference between the production rate of a top-antitop pair and the rate when top and antitop are interchanged. For observables at hadron colliders, the key quantity is the partonic charge-asymmetric cross section as a function of the top- or antitop-quark scattering angle $\theta$ with respect to the direction of the incident  parton $p_1$ in the $t\bar t$ system,\footnote{The $t\bar t$ system is the reference frame obtained by a boost of the parton CM frame along the beam axis, where the rapidities of the top and the antitop quark are of equal magnitude, but of opposite sign.}
\begin{eqnarray}\label{eq:sigmaapartdiff}
\frac{\text{d}\hat{\sigma}_{a}}{\text{d}\cos\theta} & = & \frac{\text{d}\hat{\sigma}_{t\bar t}}{\text{d}\cos\theta}\bigg\vert_{\theta = \theta_t^{t\bar t}} - \frac{\text{d}\hat{\sigma}_{\bar t t}}{\text{d}\cos\theta}\bigg\vert_{\theta = \theta_{\bar t}^{t \bar t}}\,.
\end{eqnarray}
The charge-symmetric differential cross section $\text{d}\hat{\sigma}_s/\text{d}\cos\theta$ is defined by replacing ``$-$'' with ``$+$'' in (\ref{eq:sigmaapartdiff}). After integrating over one hemisphere,\footnote{For $t\bar t + j$ production, the integration over the four remaining phase space variables is subject to a cut on the transverse momentum of the hard jet, $p_T^j$.} one obtains the total normalized partonic charge asymmetry
\begin{eqnarray}\label{eq:ac_parton}
\hat{A}_C(\hat{s}) & = & \frac{\hat{\sigma}_{a}(\hat{s})}{\hat{\sigma}_{s}(\hat{s})}\,,\qquad \hat{\sigma}_{s/a}(\hat{s}) = \int_0^1 \text{d}\cos\theta\,\frac{\text{d}\hat{\sigma}_{s/a}}{\text{d}\cos\theta}\,,
\end{eqnarray}
which depends on the partonic CM energy $\sqrt{\hat{s}}$. After convoluting with the parton distribution functions (PDFs) $f_{p_i/N_i}(x_i,\mu_f)$ at the factorization scale $\mu_f$, the hadronic charge asymmetry is given by
\begin{eqnarray}\label{eq:chargeasymmetry}
 A_C & = & \frac{\sigma_a}{\sigma_s}\,,\qquad \sigma_{s/a} = \sum_{p_1 p_2}\int \text{d} x_1 \text{d} x_2 \,f_{p_1/N_1}(x_1,\mu_f)\,f_{p_2/N_2}(x_2,\mu_f)\,\hat{\sigma}_{s/a}^{p_1 p_2}(\hat{s},\mu_f)\,.
\end{eqnarray}
The momentum fractions $x_i$ of the partons $p_i$ inside the nucleons $N_i$ determine $\hat{s}$ as a fraction of the squared CM energy of the colliding nucleons, $s$, via the relation $\hat{s} = x_1 x_2 s$. An asymmetry $\sigma_a$ is generated in the channels $p_1 p_2 = q\bar q$, $\bar q q$, $qg$, $gq$, $\bar qg$, $g\bar q$; the cross section $\sigma_s$ receives additional contributions from the charge-symmetric $gg$ initial state.

\begin{figure}[!t]
\begin{center}
\includegraphics[height=2.5cm]{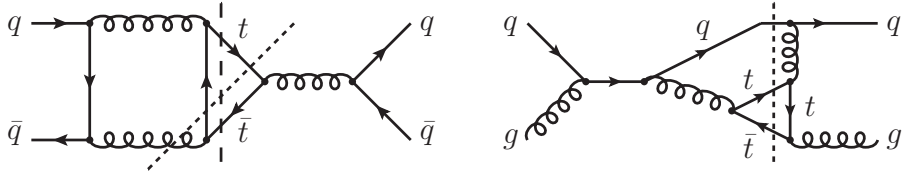}
\end{center}
\vspace*{-0.7cm}
\begin{center} 
  \parbox{15.5cm}{\caption{\label{fig:asym_qcd} Charge asymmetry in hadronic top-antitop production from quark-antiquark annihilation (left) and quark excitation (right) in QCD. Shown are representative contributions to inclusive $t\bar t$ production (dashed and dotted cuts) and exclusive $t\bar t + j$ production (dotted cuts).}}
\end{center}
\end{figure}

Since strong interactions are charge-preserving, a charge asymmetry in QCD arises only if an additional real or virtual gluon is involved in the production process of a $t\bar t$ pair. In Figure~\ref{fig:asym_qcd}, we show representative diagrams for contributions to the charge asymmetry from quark-antiquark ($q\bar q$) and quark-gluon ($qg$) initial parton states. The $qg$ contribution to the charge asymmetries both at the Tevatron and at the LHC is suppressed with respect to the dominant $q\bar q$ channel~\cite{Halzen:1987xd,Kuhn:2011ri}. In inclusive $t\bar t$ production, an asymmetry is generated at NLO by both real and virtual radiative corrections (cuts marked by dotted and dashed lines), whereas in exclusive $t\bar t + j$ production only real radiation contributes (dotted lines). The exclusive channel $t\bar t + j$ thereby allows one to disentangle the (negative) real and (positive) virtual contributions to the inclusive $t\bar t$ asymmetry by giving separate access to the subset of real contributions. In the following, we will investigate the charge asymmetry in $t\bar t + j$ production in detail. Unless stated otherwise, we will refer to $A_C$ defined in (\ref{eq:chargeasymmetry}) as the exclusive asymmetry generated by real radiation only.
%%%%%%%%%%%%%%%%%%%%%%%%%%%%%%%%%%%%%%%%%%%%%%%%%%%%%%%%%%%%%%%
%%%%%%%%%%%%%%%%%%%%%%%%%%%%%%%%%%%%%%%%%%%%%%%%%%%%%%%%%%%%%%%

At hadron colliders, observables to measure the charge asymmetry have to be constructed according to the symmetry of the initial state. At the Tevatron, the charge-asymmetric proton-antiproton initial state allows one to directly access the asymmetry $A_C$ by measuring the production rate in terms of top-antitop rapidity difference,
\begin{eqnarray}\label{eq:ACy}
A_{C}^{y} & = & \frac{\sigma(\Delta y > 0) - \sigma(\Delta y < 0)}{\sigma(\Delta y > 0) + \sigma(\Delta y < 0)} = A_C \,,\qquad \Delta y = y_{t}-y_{\bar{t}}\,.
\end{eqnarray}
The charge asymmetry is thus observed as a top-quark forward-backward asymmetry in the $t\bar t$ frame. At the LHC, one cannot probe the charge asymmetry as a forward-backward asymmetry because the proton-proton initial state is charge-symmetric. The direction of the incident quark (which sets the reference axis for the $t\bar t$ asymmetry) can still be determined by exploiting the quark's boost inside the proton. A top-quark ``beamward-central'' asymmetry can then be defined in terms of absolute rapidities\footnote{The notation $\sigma_a(\Delta|y| > 0)$ is to be understood as $\sigma_a(\Delta|y|>0) = \sigma_{t\bar t(j)}(|y_t| > |y_{\bar t}|) - \sigma_{\bar t t(j)}(|y_{\bar t}| > |y_{t}|)$.}
\begin{eqnarray}\label{eq:ACabsy}
A_{C}^{|y|} & = & \frac{\sigma(\Delta |y| > 0) - \sigma(\Delta |y| < 0)}{\sigma(\Delta |y| > 0) + \sigma(\Delta |y| < 0)} = \frac{\sigma_a(\Delta|y|>0)}{\sigma_s} < A_C\,,\qquad \Delta|y|=|y_{t}|-|y_{\bar{t}}|\,.
\end{eqnarray}
This observable, however, is much smaller than the Tevatron asymmetry $A_C^y$ for two reasons: Firstly, $A_C^{|y|}$ is defined in the laboratory (lab) frame and is therefore diluted by the boost with respect to the $t\bar t$ frame, resulting in $A_C^{|y|} < A_C$. Secondly and mainly, the charge-symmetric $gg$ background at the LHC is much larger than at the Tevatron, yielding $A_C^{(\text{LHC})}\ll A_C^{(\text{Tevatron})}$.

\subsection{Phase space variables and cuts}\label{sec:phasespace}
The strong suppression of the total charge asymmetry at the LHC can be relaxed by applying suitable cuts on the final-state phase space. In particular, the jet kinematics can be used to enhance the asymmetry in $t\bar t + j$ production. In this section, we define a set of cuts in terms of phase space variables that allows one to systematically probe the kinematic behavior of the cross section $\sigma_s$ and the charge asymmetries $A_C^y$ and $A_C^{|y|}$ at the Tevatron and the LHC.

The phase space of the partonic process $p_1 p_2\rightarrow t\bar t j$ can be expressed in terms of five variables in the $p_1 p_2$ CM frame. We choose these variables to be the jet and top-quark energies, $E_j$ and $E_t$, the jet and top-quark scattering angles with respect to the incident parton $p_1$, $\theta_j$ and $\theta_t$, and the azimuthal angle of the top quark, $\phi_t$. Hadronic observables depend on two further variables: the boost of the partonic CM frame with respect to the lab frame, $\beta = (x_1-x_2)/(x_1+x_2)$, and the squared CM energy of the $p_1p_2$ pair, $\hat{s}$. Both these variables are fixed by the momentum fractions $x_1$ and $x_2$ carried by the partons inside the colliding hadrons. 

The top-quark energy $E_t$ and the scattering angle $\theta_t$, as well as their counterparts $E_{\bar t}$ and $\theta_{\bar t}$ for the antitop quark, determine the rapidities in the $p_1 p_2$ CM frame,
\begin{eqnarray}
 \hat{y}_{t} = \frac{1}{2}\ln\left(\frac{1 + \sqrt{1-m_t^2/E_t^2}\,\cos\theta_{t}}{1 - \sqrt{1-m_t^2/E_t^2}\,\cos\theta_{t}}\right)\quad \text{and} \quad \hat{y}_{\bar t} = \frac{1}{2}\ln\left(\frac{1 + \sqrt{1-m_t^2/E_{\bar t}^2}\,\cos\theta_{\bar t}}{1 - \sqrt{1-m_t^2/E_{\bar t}^2}\,\cos\theta_{\bar t}}\right)\,.
\end{eqnarray}
The rapidities that define the charge asymmetries in (\ref{eq:ACy}) and (\ref{eq:ACabsy}) are obtained after boosting $\hat{y}_t$ and $\hat{y}_{\bar t}$ to the lab frame via $y_{t/\bar t} = \hat{y}_{t/\bar t} + y_{t\bar tj}$, with $y_{t\bar tj}$ defined in Appendix~\ref{app:phasespace}. As the charge asymmetry increases with the rapidity difference $\Delta y$ ($\Delta |y|$), a lower cut on this quantity helps to enhance $A_C^y$ ($A_C^{|y|}$). This cut does not involve the jet kinematics and is thus equally applicable in inclusive $t\bar t$ and exclusive $t\bar t + j$ production.

Jet kinematics are parameterized by the jet energy $E_j$ and the scattering angle $\theta_j$. The production cross section $\sigma_s$ is particularly sensitive to both variables, due to the logarithmic enhancement in the region where $E_j\rightarrow 0$ and/or $\theta_j\rightarrow 0$. We postpone a detailed discussion of this behavior to Section~\ref{sec:qqbttg} and limit ourselves here to expressing $E_j$ and $\theta_j$ in terms of lab-frame quantities that are suitable for experimental cuts. At LO, the jet energy $E_j$ is related to the (frame-independent) invariant mass $M_{t\bar t}$ of the top-antitop pair via
\begin{eqnarray}\label{eq:mtt}
M_{t\bar t}^2 & = & (p_t + p_{\bar t})^2 \stackrel{\text{LO}}{=} \hat{s}\,(1-2E_j/\sqrt{\hat{s}})\,,
\end{eqnarray}
where $p_t$ and $p_{\bar t}$ denote the four-momenta of the top and antitop quarks. A lower cut on $M_{t\bar t}$ implies an upper bound on the jet energy $E_j$. The dependence of the asymmetry on $M_{t\bar t}$ is mild in QCD, but becomes important in the search for massive resonances. We will elaborate on this point in Section~\ref{sec:asymmetry-coloroctets}.

The jet scattering angle $\theta_j$ encodes important information about the asymmetry in the collinear limit $\theta_j\rightarrow 0$. The jet rapidity in the partonic CM frame, $\hat{y}_j$, is directly related to the angle $\theta_j$. For experimental purposes, it is convenient to express rapidities in the parton frame in terms of rapidity differences that can be measured in the lab frame, the latter being invariant under boosts along the beam axis. Here we exploit the feature that the rapidity of the system of final-state particles in the parton frame vanishes, which at LO implies $\hat{y}_{t\bar t j} = 0$.\footnote{The definition of $\hat{y}_{t\bar tj}$ is given as for $y_{t\bar t j}$ in Appendix~\ref{app:phasespace} by interpreting the momenta as given in the parton frame.} The parton-frame rapidity $\hat{y}_j$ can then be described conveniently in terms of the lab-frame rapidities $y_j$ and $y_{t\bar tj}$ as
\begin{eqnarray}\label{eq:jet_rapidity}
\hat{y}_j & = & \frac{1}{2}\ln\left(\frac{1+\cos\theta_j}{1-\cos\theta_j}\right) \stackrel{\text{LO}}{=} \hat{y}_j - \hat{y}_{t\bar tj} = y_j - y_{t\bar tj}\,.
\end{eqnarray}
As we will show in Section~\ref{sec:qqbttg}, the QCD charge asymmetry is maximal if the jet is emitted perpendicular to the beam axis. A cut on the jet rapidity $\hat{y}_j$ is therefore suitable to enhance the integrated charge asymmetry in QCD.

The azimuthal angle of the top quark $\phi_t$ is not observable at hadron colliders because the experimental setup is symmetric under rotations around the beam axis. What can be observed is the azimuthal correlation between the particles in the final state, for instance the difference between the azimuthal angles of the top quark and the antitop quark, $\Delta\phi_{t\bar t} = \phi_t - \phi_{\bar t}$. For fixed transverse momenta $p_T^t = E_t\sin\theta_t$ and $p_T^{\bar t} = E_{\bar t}\sin\theta_{\bar t}$, this azimuthal difference can be expressed in terms of the jet transverse momentum $p_T^j = E_j\sin\theta_j$ via
\begin{eqnarray}
(p_T^j)^2 & \stackrel{\text{LO}}{=} (p_T^{t\bar t})^2 = (p_T^t)^2 + (p_T^{\bar t})^2 + 2\,p_T^t p_T^{\bar t} \cos\Delta\phi_{t\bar t}\,.
\end{eqnarray}
The dependence of the charge asymmetry on $p_T^j$ will be discussed in Section~\ref{sec:jetcuts}. Azimuthal correlation in terms of $\Delta \phi_{t\bar t}$ will not be a subject of this work.

The boost $\beta$ of the partonic CM frame with respect to the lab frame helps to discriminate between different partonic initial states \cite{AguilarSaavedra:2011cp}. It is related to the lab-frame rapidity of the $t\bar t + j$ system, $y_{t\bar tj}$, via 
\begin{eqnarray}\label{eq:yttbj}
y_{t\bar tj} & \stackrel{\text{LO}}{=} & \frac{1}{2}\ln\left(\frac{1+\beta}{1-\beta}\right)\,,\qquad \text{with}\quad \beta = \frac{x_1-x_2}{x_1+x_2}\,.
\end{eqnarray}
The derivation, as well as an illustration of the relation between $\beta$ and $y_{t\bar tj}$ are given in Appendix \ref{app:phasespace}. Accordingly, a lower cut on $|y_{t\bar tj}|$ selects events with a strong boost $\beta$ along the beam line. It increases the weight of the partonic states with a boosted valence quark, i.e. $q\bar q$ and $qg$, and thereby enhances the charge asymmetry over its $gg$ background. This feature is thus of particular importance at the LHC, where the fraction of $gg$ initial states dominates the $t\bar t + j$ production cross section.\\

Besides the above-mentioned ``strategic'' cuts, the design of the detector and the methods of jet reconstruction require additional cuts on the jet properties. If not stated otherwise, we work with the following cuts on the jet rapidity $y_j$ and the transverse momentum $p_T^j$. For Tevatron observables, we employ $|y_j| \le 2$ and $p_T^j \ge 20\,\text{GeV}$, for the LHC we use $|y_j| \le 2.5$ and $p_T^j \ge 25\,\text{GeV}$. The rapidity cut prevents the jet from being emitted parallel to the beam axis, enforcing $\theta_j\neq 0$, whereas the transverse momentum cut imposes a minimum jet energy $E_j > 0$ for a given jet direction. For our numerical analysis, we set the factorization and renormalization scales equal to the top-quark mass $\mu_f = \mu_r = m_t = 173.2\,\text{GeV}$ \cite{Lancaster:2011wr}. Since all calculations are done at LO, we use CTEQ6L1 PDFs \cite{Pumplin:2002vw} with an input of the strong coupling constant $\alpha_s(m_t) = 0.1180$. The phase-space integration is performed numerically by means of the Vegas Monte-Carlo algorithm implemented in \cite{Hahn:2004fe,Galassi}.\\

\subsection{Anatomy of $t\bar t + j$ production at parton level: the jet handle}\label{sec:qqbttg}
\begin{figure}[!t]
\begin{center}
\includegraphics[height=5cm]{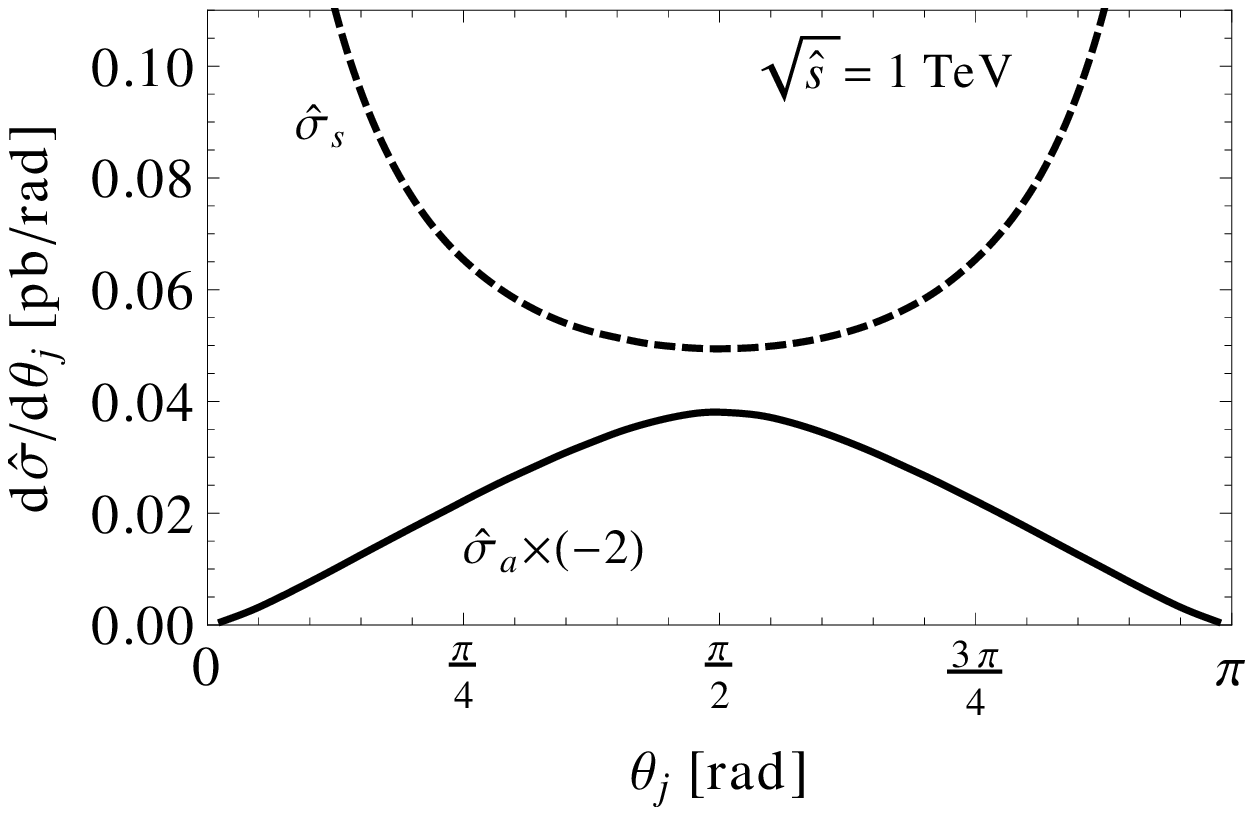}
\hspace*{1.5cm}
\raisebox{2.57cm}{
\begin{tabular}{|c c|c|c|}
\hline
 & & soft & collinear\tabularnewline
\hline
\hline
$q\bar q$: & $\hat{\sigma}_s$ & \checkmark & \checkmark \tabularnewline
 & $\hat{\sigma}_a$ & \checkmark & -- \tabularnewline
\hline
$qg$: & $\hat{\sigma}_s$ & -- & \checkmark \tabularnewline
 & $\hat{\sigma}_a$ & -- & -- \tabularnewline
\hline
$gg$: & $\hat{\sigma}_s$ & \checkmark & \checkmark \tabularnewline
\hline
\end{tabular}}
\end{center}
\vspace*{-0.8cm}
\begin{center} 
  \parbox{15.5cm}{\caption{\label{fig:jet_dist_parton} Left: Jet angular distribution in QCD for the partonic process $q\bar q \rightarrow t\bar t g$ as a function of the scattering angle $\theta_j$ for $\sqrt{\hat{s}} = 1\,\text{TeV}$ and $p_T^{j}\ge 2\,\text{GeV}$. The charge-asymmetric cross section $\hat{\sigma}_a$ (solid line) is rescaled by a factor of $-2$. Right: Soft and collinear singularities of the charge-symmetric ($\hat{\sigma}_s$) and -asymmetric ($\hat{\sigma}_a$) contributions to $t\bar t + j$ production from $q\bar q$, $qg$, and $gg$ initial states.}}
\end{center}
\end{figure}

Before turning to measurable quantities, we discuss in detail the kinematics of the emitted gluon in the partonic process $q\bar q\rightarrow t\bar t g$, which yields the dominant contribution to the top-quark charge asymmetry. At the parton level, the features of the cross section and the asymmetry are not diluted by boosts along the beam axis or by the ignorance of the quark direction. Due to the occurrence of singularities in the infrared limit, a good understanding of the kinematics is crucial in the region where the emitted gluon is soft, $E_j\rightarrow 0$, or (anti-) collinear to the incident quark, $\theta_j\rightarrow \{0,\pi\}$. The infrared behavior of the charge-symmetric and -asymmetric cross section for different partonic initial states is summarized in the table in Figure~\ref{fig:jet_dist_parton}. In the $q\bar q$ and $gg$ channels, the charge-symmetric cross section $\hat{\sigma}_s$ exhibits both soft and collinear singularities. The charge-asymmetric cross section $\hat{\sigma}_a$, arising from the diagrams shown in Figure~\ref{fig:asym_qcd}, is finite in the collinear limit $\theta_j\rightarrow \{0,\pi\}$. The differing collinear behavior of $\hat{\sigma}_s$ and $\hat{\sigma}_a$ is reflected by the angular distribution of the gluon,
\begin{eqnarray}\label{eq:dsigmadtheta}
\frac{\text{d}\hat{\sigma}_{s/a}}{\text{d}\theta_j} & = & \frac{\text{d}\hat{\sigma}_{s/a}(\cos\theta_t^{t\bar t} > 0)}{\text{d}\theta_j} \pm \frac{\text{d}\hat{\sigma}_{s/a}(\cos\theta_{\bar t}^{t\bar t} > 0)}{\text{d}\theta_j}\,,
\end{eqnarray}
which is displayed in Figure~\ref{fig:jet_dist_parton} for the partonic process $q\bar q\rightarrow t\bar t g$. The symmetric cross section $\hat{\sigma}_s$ (dashed line) is enhanced if the gluon is emitted along the beam line, which is due to the collinear singularity in the limit $\theta_j\rightarrow \{0,\pi\}$. The asymmetric cross section $\hat{\sigma}_a$ (solid line) does not exhibit this collinear enhancement and reaches its maximum if the gluon is emitted perpendicular to the beam line. Since QCD is invariant under CP conjugation, the angular distribution of the emitted parton from the (CP-symmetric) $q\bar q$ and $gg$ initial states is symmetric under $\theta_j\leftrightarrow \pi-\theta_j$ for both $\hat{\sigma}_s$ and $\hat{\sigma}_a$.
%%%%%%%%%%%%%%%%%%%%%%%%%%%%%%%%%%%%%%%%%%%%%%%%%%%%%%%%%%%%%%%%%%%%%%%
%%%%%%%%%%%%%%%%%%%%%%%%%%%%%%%%%%%%%%%%%%%%%%%%%%%%%%%%%%%%%%%%%%%%%%%

\begin{figure}[!t]
\begin{center}
\includegraphics[height=5cm]{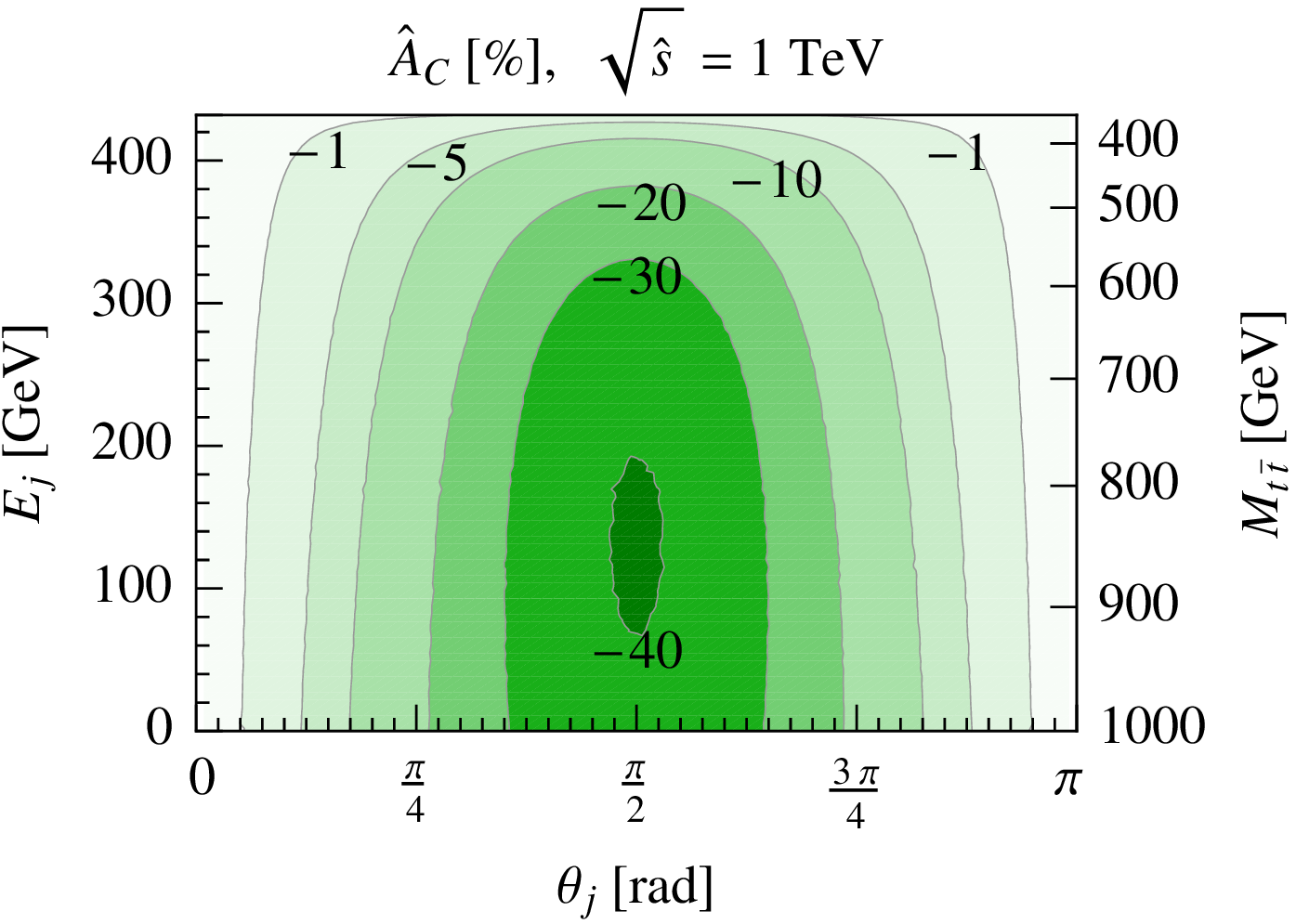}
\hspace*{1.5cm}
\includegraphics[height=5cm]{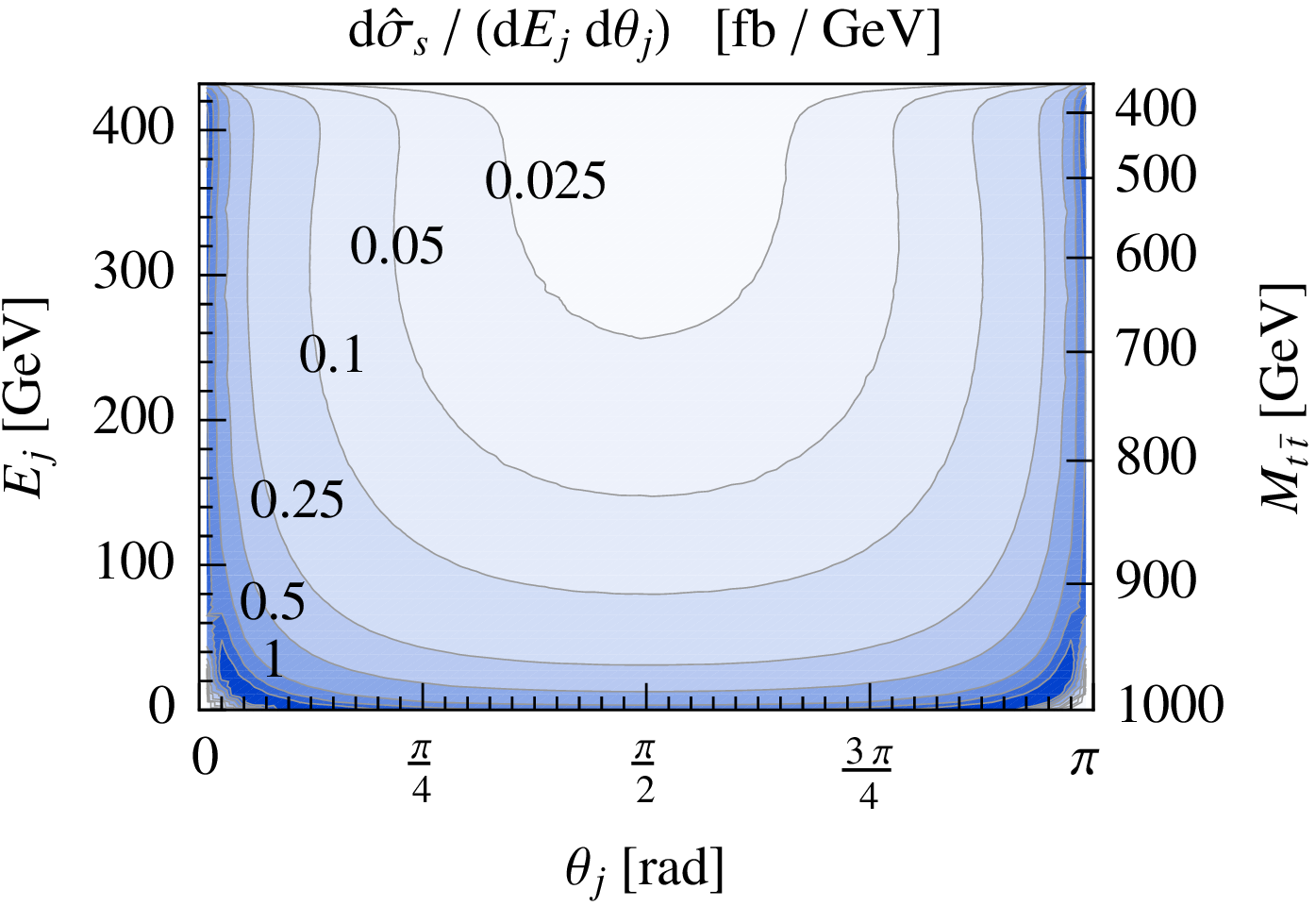}
\end{center}
\vspace*{-1cm}
\begin{center} 
  \parbox{15.5cm}{\caption{\label{fig:infrared_behavior} The partonic process $q\bar q\rightarrow t\bar t g$ for $\sqrt{\hat{s}}=1\,\text{TeV}$ and $p_T^j\ge 2\,\text{GeV}$ as a function of the gluon scattering angle $\theta_j$, the gluon energy $E_j$, and the invariant mass $M_{t\bar t}$. Left: Partonic charge asymmetry $\hat{A}_C\,[\%]$. Right: Charge-symmetric cross section $\text{d}\hat{\sigma}_s/(\text{d}E_j\,\text{d}\theta_j)\,[\text{fb} / \text{GeV}]$.}}
\end{center}
\end{figure}

A cut on the transverse momentum of the jet, $p_T^j = E_j \sin\theta_j$, simultaneously regularizes soft and collinear divergences. In the region of a low transverse momentum, QCD predicts a double-logarithmic enhancement of the charge-symmetric total cross section $\sigma_s$ and a single-logarithmic enhancement of the asymmetric cross section $\sigma_a$ \cite{Melnikov:2010iu},
\begin{equation}\label{eq:sigmadiv}
\sigma_s(p_T^{j})\sim \ln\left(\frac{m_t}{p_T^{j}}\right)_{\hspace*{-0.15cm}\text{soft}}\hspace*{-0.38cm}\cdot\ln\left(\frac{m_t}{p_T^{j}}\right)_{\hspace*{-0.15cm}\text{coll.}},\qquad \sigma_a(p_T^{j})\sim \ln\left(\frac{m_t}{p_T^{j}}\right)_{\hspace*{-0.15cm}\text{soft}}.
\end{equation}
The interplay of collinear and soft divergences is shown in Figure~\ref{fig:infrared_behavior} for the parton channel $q\bar q\rightarrow t\bar t g$. In this channel, the partonic cross sections $\hat{\sigma}_s$ and $\hat{\sigma}_a$ feature the same infrared behavior as the hadronic quantities $\sigma_s$ and $\sigma_a$ in (\ref{eq:sigmadiv}). We display the partonic charge asymmetry $\hat{A}_C$ defined in (\ref{eq:ac_parton}) (left panel) and the cross section $\hat{\sigma}_s$ (right panel) as a function of the gluon scattering angle $\theta_j$ and the gluon energy $E_j$, the latter being connected to the invariant mass $M_{t\bar t}$ via (\ref{eq:mtt}). The double-logarithmic enhancement of $\hat{\sigma}_s$ from~(\ref{eq:sigmadiv}) is visible in the regions where $E_j\rightarrow 0$ and $\theta_j\rightarrow \{0,\pi\}$. The charge asymmetry $\hat{A}_C$ shows a mild dependence on $E_j$ or $M_{t\bar t}$, unless $E_j$ ($M_{t\bar t}$) is very large (small). In the soft-gluon limit, where most of the CM energy $\sqrt{\hat{s}}$ is carried by the $t\bar t$ pair, the logarithmic enhancement of $\hat{\sigma}_s$ and $\hat{\sigma}_a$ from (\ref{eq:sigmadiv}) cancels in the ratio $\hat{A}_C$. The normalized asymmetry is thus largely unaffected by a variation of the gluon energy $E_j$ in the soft region. The low sensitivity to $E_j$ or $M_{t\bar t}$ translates to the hadronic asymmetries $A_C^y$ and $A_C^{|y|}$, as the $t\bar t + j$ cross section $\sigma_s$ is dominated by soft jets (cf. Figure~\ref{fig:infrared_behavior}, right). Still, a lower cut on $M_{t\bar t}$ increases the asymmetry, as it projects on the region of soft-gluon emission even for larger CM energies, which yields the dominant contribution. The dependence of the charge asymmetry on $\theta_j$, in turn, is strong. Due to the collinear enhancement of $\hat{\sigma}_s$, but not of $\hat{\sigma}_a$, the charge asymmetry $\hat{A}_C \sim \ln^{-1}(m_t/p_T^{j})$ vanishes in the limit $\theta_j\rightarrow \{0,\pi\}$. The maximum of $\hat{A}_C$ is reached for $\theta_j=\pi/2$ due to the synergy of two effects: Firstly, $\hat{\sigma}_a$ is maximal for a gluon emitted perpendicular to the beam axis and secondly, $\hat{\sigma}_s$ entering the normalization is depleted in the central region (cf. Figure~\ref{fig:jet_dist_parton}, left). A lower cut on the gluon scattering angle $\theta_j$, or equivalently on the jet rapidity $\hat{y}_j$ from (\ref{eq:jet_rapidity}) for hadronic observables, therefore increases the charge asymmetry. In addition, it has the beneficial feature of focusing on the collinear-safe region, where the asymmetry is not affected by logarithmic divergences. In the collinear limit $\theta_j\rightarrow \{0,\pi\}$, in turn, large logarithms should be resummed in order to make a reliable prediction of the charge asymmetry in this region.
%%%%%%%%%%%%%%%%%%%%%%%%%%%%%%%%%%%%%%%%%%%%%%%%%%%%%%%%%%%%%%%%%%%%%%%
%%%%%%%%%%%%%%%%%%%%%%%%%%%%%%%%%%%%%%%%%%%%%%%%%%%%%%%%%%%%%%%%%%%%%%%

\subsection{Forward-backward asymmetry at the Tevatron}\label{sec:afbtevatron}
Top-quark pair production at the Tevatron proceeds mainly through the partonic channel $q\bar q$. This feature is due to the CM energy of $\sqrt{s}=1.96\,\text{TeV}$, where the $gg$ channel is suppressed by the PDFs, and to the fact that both $q$ and $\bar q$ are valence quarks in the initial hadronic state $p\bar p$. The total cross section for the process $p\bar p\rightarrow t\bar t + j$ in QCD at LO is $\sigma_s^{\text{LO}} = 1.63^{+0.99}_{-0.54}\,\text{pb}$ for $p_T^j\ge 20\,\text{GeV}$,\footnote{The uncertainties on $\sigma_s^{\text{LO}}$ and $A_C^{y,\text{LO}}$ (as well as on $\sigma_s^{\text{LO}}$ and $A_C^{|y|,\text{LO}}$ in Section~\ref{sec:abclhc}) are due to the scale variations $m_t/2 < \mu_f=\mu_r < 2m_t$. No detector cut on $|y_j|$ has been applied in order to allow a comparison with the NLO results from \cite{Dittmaier:2008uj} and \cite{Alioli:2011as}.} where the respective contributions of the partonic channels amount to $83\%$ ($q\bar{q}$), $10\%$ ($qg+\bar{q}g$), and $7\%$ ($gg$). At NLO, the cross section is given by $\sigma_s^{\text{NLO}} = 1.791(1)^{+0.16}_{-0.31}\,\text{pb}$ \cite{Dittmaier:2008uj},\footnote{The results for $\sigma_s^{\text{NLO}}$ and $A_C^{y,\text{NLO}}$ are obtained with $m_t = 174\,\text{GeV}$ and CTEQ6M PDFs.} which constitutes about $26\,\%$ of the total cross section for inclusive $t\bar t$ production \cite{Ahrens:2011px}. The QCD prediction of the total cross section agrees with the measurement by the CDF collaboration, $\sigma_s = 1.6\pm 0.2\pm 0.5\,\text{pb}$ \cite{CDF:2009}. The integrated charge asymmetry defined in (\ref{eq:ACy}) amounts to $A_C^{y,\text{LO}} = -11.1^{+0.2}_{-0.1}\,\%$ at LO. It is affected by large NLO QCD corrections, yielding $A_C^{y,\text{NLO}} = -4.40\pm 0.04\,\%$ \cite{Alioli:2011as}. This large reduction is of the order of the charge asymmetry in inclusive $t\bar t$ production, $A_C^{y,t\bar t} \approx 7\,\%$. It does not signal a breakdown of the perturbative framework, but is due to contributions from virtual hard-gluon corrections \cite{Melnikov:2010iu}.

\begin{figure}[!t]
\begin{center}
\includegraphics[height=5cm]{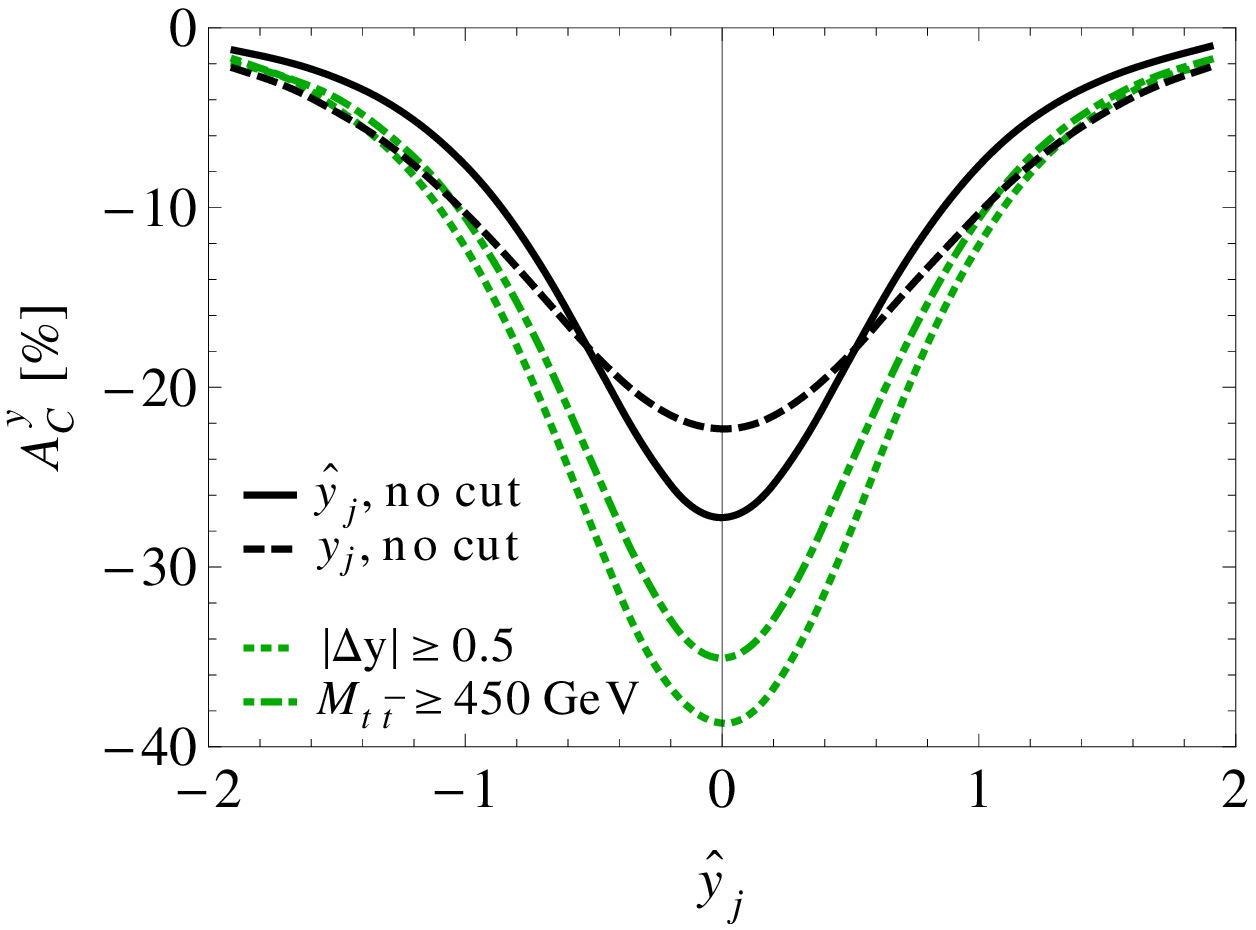}
\hspace*{1.5cm}
\includegraphics[height=5cm]{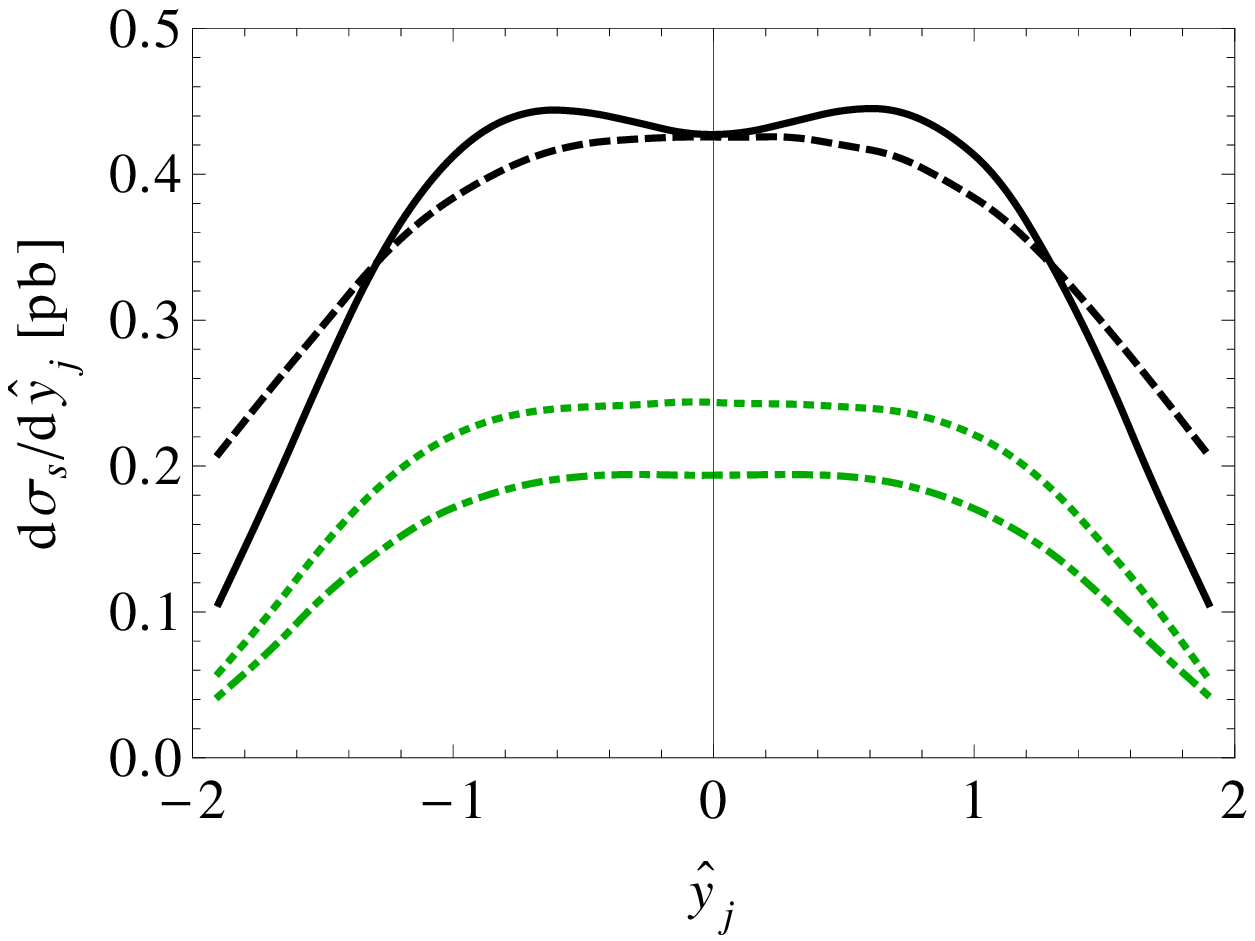}
\end{center}
\vspace*{-1cm}
\begin{center} 
  \parbox{15.5cm}{\caption{\label{fig:tev_ac_sigmas} Charge asymmetry $A_C^y$ (left) and differential cross section $\text{d}\sigma_s/d\hat{y}_j$ (right) for $t\bar t + j$ production at the Tevatron as a function of the jet rapidity in the parton CM frame $\hat{y}_j$ (black) and the lab frame $y_j$ (dashed black). Additional cuts on the $t\bar t$ rapidity difference $\Delta y$ (invariant mass $M_{t\bar t}$) are shown for the $\hat{y}_j$ distributions in dotted green/gray (dot-dashed green/gray). Detector cuts of $|y_j| \le 2$ and $p_T^j \ge 20\,\text{GeV}$ have been applied for all distributions.}}
\end{center}
\end{figure}

In the following, we discuss the dependence of the Tevatron asymmetry on the jet kinematics. The low $gg$ background and the equivalence of the charge asymmetry $A_C$ and the forward-backward asymmetry $A_C^y$ from (\ref{eq:ACy}) allow us to apply our findings for the partonic channel $q\bar q\rightarrow t\bar t g$ from Section~\ref{sec:qqbttg} directly to the hadronic observables. We saw that the asymmetry is strongly dependent on the jet rapidity. In the left panel in Figure~\ref{fig:tev_ac_sigmas}, we show the charge asymmetry $A_C^y$ as a function of the jet rapidity in the partonic CM frame $\hat{y}_j$ (solid black curve), compared to the lab-frame rapidity $y_j$ (dashed black curve). The difference between $\hat{y}_j$ and $y_j$ is given by the boost of the parton frame encoded in the rapidity $y_{t\bar tj}$, as has been shown in (\ref{eq:jet_rapidity}). Due to this boost, the distribution in terms of the lab-frame rapidity $y_j$ is washed out with respect to the parton-frame rapidity $\hat{y}_j$. The latter distribution reaches its maximum if the jet is emitted perpendicular to the beam axis, yielding $A_C^y(\hat{y}_j=0) = -27\%$. An upper cut on $|\hat{y}_j|$ is therefore well suited to enhance the asymmetry, though at the cost of reducing the cross section $\sigma_s$ displayed on the right-hand side in Figure~\ref{fig:tev_ac_sigmas}. Notice that due to the significant dependence on $y_j$, the integrated charge asymmetry is very sensitive to the detector cut $|y_j| \le 2$.

%%%%%%%%%%%%%%%%%%%%%%%%%%%%%%%%%%%%%%%%%%%%%%%%%%%%%%%%%%%%%
%%%%%%%%%%%%%%%%%%%%%%%%%%%%%%%%%%%%%%%%%%%%%%%%%%%%%%%%%%%%%

As discussed in Section~\ref{sec:qqbttg}, the asymmetry can also be enhanced by performing cuts on the invariant mass $M_{t\bar{t}}$, related to the jet energy $E_j$ via (\ref{eq:mtt}), and on the top-antitop rapidity difference $\Delta y$. In Table~\ref{tab:tev_cuts}, we display the Tevatron charge asymmetry $A_C^y$ and the cross section $\sigma_s$ for cuts on $|\hat{y}_j|$, $M_{t\bar t}$ and $|\Delta y|$. This sample allows us to compare the various cuts with respect to their efficiency to enhance the charge asymmetry, while keeping the reduction of the cross section moderate. The second, fourth and fifth columns of the table show that a cut on $|\hat{y}_j|$ can enhance $A_C^y$ from $-12.6\,\%$ to the $-20\%$ level, but simultaneously reduces $\sigma_s$ by about a factor of two or more. Comparing with the third column, one observes that a cut on $M_{t\bar t}$ is less efficient than $|\hat{y}_j|$ in enhancing the charge asymmetry for a given cross section. The efficiency of cuts on $|\hat{y}_j|$ and $|\Delta y|$ (columns four and six) is almost the same for cross sections down to $\sigma_s = 0.4\,\text{pb}$, or equivalently for $|\hat{y}_j|^{\text{max}}\ge 0.5$ and $|\Delta y|^{\text{min}} \le 0.8$. By comparing the fifth and the last columns, it is apparent that a combination of cuts on $|\hat{y}_j|$ and $|\Delta y|$ is more efficient to enhance the asymmetry for a fixed cross section than a stronger cut on $|\hat{y}_j|$ alone. The effect of additional cuts on $M_{t\bar t}$ and $|\Delta y|$ is illustrated in Figure~\ref{fig:tev_ac_sigmas}, where $A_C^y(\hat{y}_j)$ and $\text{d}\sigma_s/\text{d}\hat{y}_j$ are displayed for $|\Delta y| \ge 0.5$ (dotted green/gray curves) and $M_{t\bar t}\ge 450\,\text{GeV}$ (dot-dashed green/gray curves). Due to the low $t\bar t + j$ production rate at the Tevatron, however, a measurement of the charge asymmetry and its kinematical features remains an experimental challenge.

%%%%%%%%%%%%%%%%%%%%%%%%%%%%%%%%%%%%%%%%%%%%%%%%%%%%%%%%%%%%%
\begin{table}[!b]
\begin{center}
\begin{tabular}{|c|c|c|c|c|c|c|}
\hline 
 & no cuts & $M_{t\bar{t}}\ge 450\,\text{GeV}$ & $|\hat{y}_{j}|\le 1$ & $|\hat{y}_{j}|\le 0.5$ & $|\Delta y| \ge 0.5$ & $|\hat{y}_{j}| \le 1$, $|\Delta y| \ge 0.5$\tabularnewline
\hline
\hline
$A_{C}^y\,\,[\%]$ & $-12.6$ & $-17.0$ & $-18.2$ & $-24.0$ & $-19.1$ & $-27.5$\tabularnewline
\hline 
$\sigma_{s}\,\,[\text{pb}]$ & $1.42$ & $0.61$ & $0.87$ & $0.43$ & $0.78$ & $0.48$\tabularnewline
\hline
\end{tabular}
\end{center}
\begin{center} 
  \parbox{15.5cm}{\caption{\label{tab:tev_cuts} Charge asymmetry $A_C^y$ and cross section $\sigma_s$ at the Tevatron for cuts on the invariant mass $M_{t\bar t}$, the partonic jet rapidity $\hat{y}_j$, and the rapidity difference $\Delta y = y_t - y_{\bar t}$.}}
\end{center}
\end{table}
%%%%%%%%%%%%%%%%%%%%%%%%%%%%%%%%%%%%%%%%%%%%%%%%%%%%%%%%%%%%%

The charge asymmetry in $t\bar t + j$ production has been investigated before in \cite{Alioli:2011as} at NLO QCD including parton showering. The enhancement of $A_C^y$ for the applied cuts $M_{t\bar t} \ge 450\,\text{GeV}$ and $|\Delta y| \ge 1.0$ qualitatively agrees with our findings at LO. Numerically, however, the NLO results for $A_C^y$ differ significantly from our LO results due to the large shift of the asymmetry by NLO corrections. In order to obtain precise QCD predictions of the charge asymmetry for the cuts suggested in this work, an analysis of the kinematic distributions at NLO would be necessary.

\subsection{Beamward-central asymmetry at the LHC}\label{sec:abclhc}
The difficulty in observing a hard jet in $t\bar t$ production at the Tevatron is overcome at the LHC by a much larger production rate. The QCD observables for the process $pp\rightarrow t\bar t + j$ have been previously calculated at NLO for a CM energy of $\sqrt{s} = 7\,\text{TeV}$ and a cut on the jet transverse momentum $p_T^j \ge 50\,\text{GeV}$. At LO, the total cross section is given by $\sigma_s^{\text{LO}} = 40.9^{+23.6}_{-13.2}\,\text{pb}$. This is to be compared with the cross section at NLO, $\sigma_s^{\text{NLO}} = 53.1(2)^{+4.1}_{-8.9}\,\text{pb}$ \cite{Alioli:2011as},\footnote{The result for $\sigma_s^{\text{NLO}}$ includes effects from parton showering.} which makes up $29\,\%$ of the inclusive $t\bar t$ cross section at NLO given in \cite{Ahrens:2011px}. The fraction of $t\bar t + j$ events in the inclusive $t\bar t$ sample increases when lowering the cut on $p_T^j$. Compared to $t\bar t$ production at the Tevatron, the contribution of $t\bar t + j$ is larger due to the enlarged phase space for the production of a hard jet at higher collision energies. The charge asymmetry defined in (\ref{eq:ACabsy}) is given at LO by $A_C^{|y|,\text{LO}} = -0.47\pm 0.04\,\%$. The large contribution of virtual gluon corrections at NLO causes a sign shift, yielding $A_C^{|y|,\text{NLO}} = 0.51 \pm 0.09\,\%$ \cite{Alioli:2011as}.

We will perform our analysis of the charge asymmetry at the LHC for $\sqrt{s} = 8\,\text{TeV}$ (LHC8), $p_T^j \ge 25\,\text{GeV}$, and a detector cut of $|y_j| \le 2.5$. The total cross section at LO is given by $\sigma_s^{\text{LO}} = 97.5^{+54.0}_{-30.6}\,\text{pb}$. Compared to the result for $\sqrt{s} = 7\,\text{TeV}$ and $p_T^j \ge 50\,\text{GeV}$, the cross section is enhanced by the increased CM energy, but mainly due to the relaxed cut on the transverse momentum of the jet. The respective partonic channels contribute to the cross section with $7.7\%$ ($q\bar{q}$), $26.7\%$ ($qg + \bar{q}g$) and $65.6\%$ ($gg$). The charge asymmetry is given by $A_C^{|y|,\text{LO}} = -0.56\pm 0.5\,\%$. As was explained previously, the strong suppression of the charge asymmetry at the LHC is mainly due to the large charge-symmetric $gg$ background and to the fact that the asymmetry in the $qg$ channel is tiny. Thus, in order to observe a sizeable charge asymmetry in $t\bar t + j$ production at the LHC, the application of cuts is indispensable.

\begin{figure}[!t]
\begin{center}
\includegraphics[height=5cm]{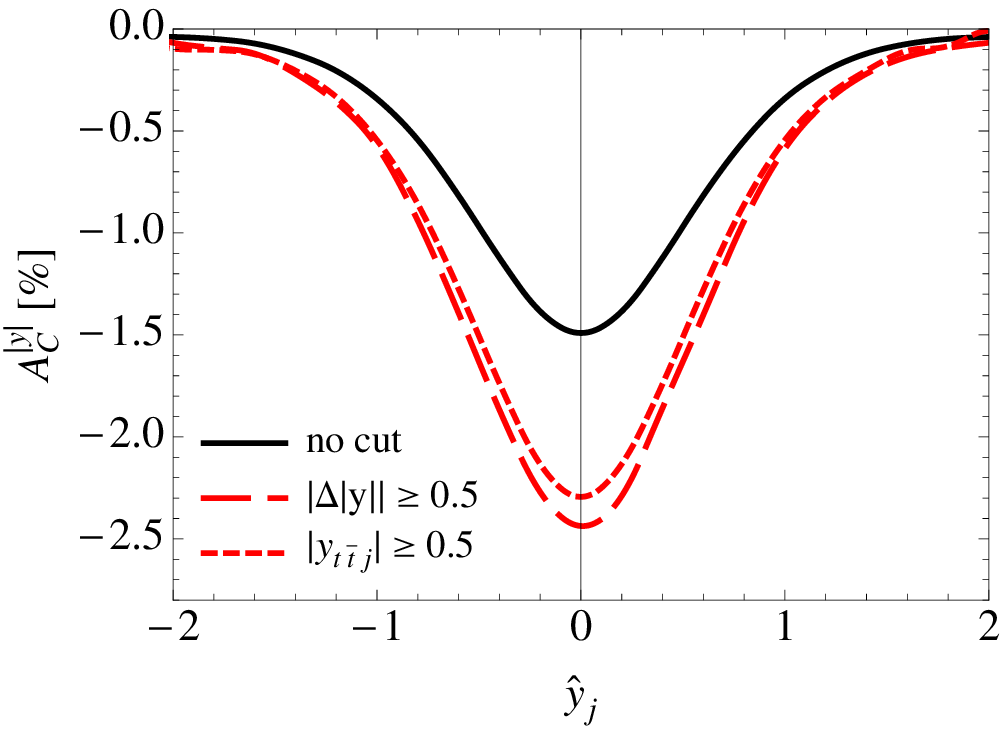}
\hspace*{1.5cm}
\includegraphics[height=5cm]{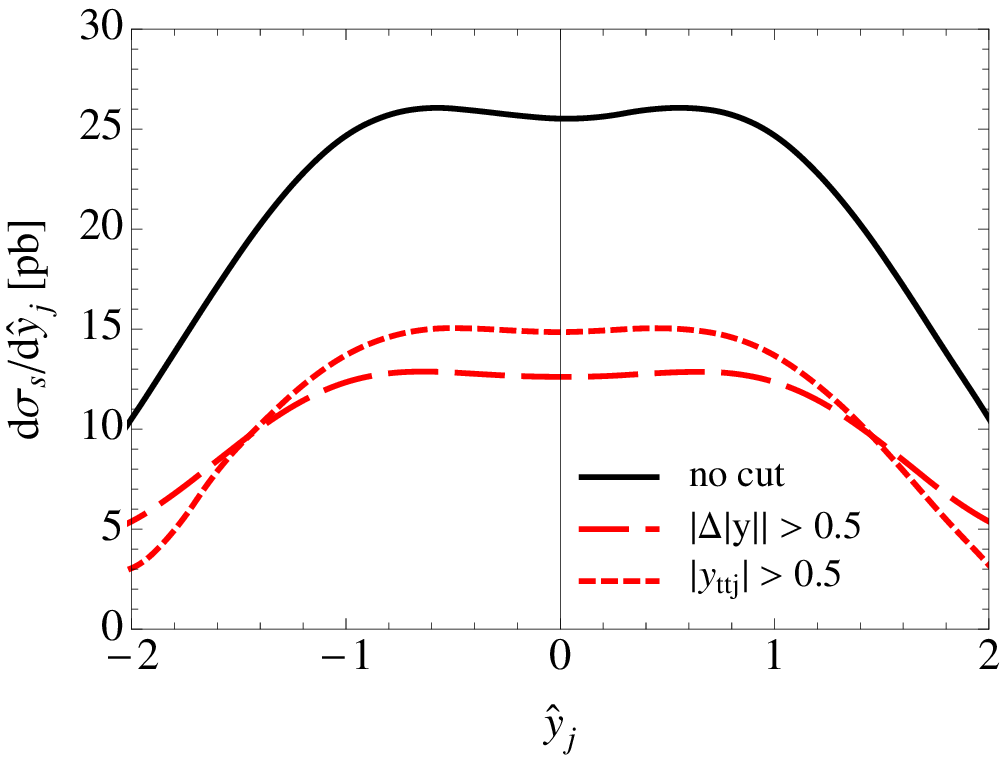}
\end{center}
\vspace*{-1cm}
\begin{center} 
  \parbox{15.5cm}{\caption{\label{fig:lhc_ac_sigmas} Charge asymmetry $A_C^{|y|}$ (left) and cross section $\sigma_s$ (right) for $t\bar t + j$ production at LHC8 as a function of the parton-frame rapidity $\hat{y}_{j}$ without (black) and with additional cuts on the parton-frame boost $|y_{t\bar tj}|$ (dashed red/gray) and the difference between absolute $t$ and $\bar t$ rapidities $|\Delta |y||$ (long-dashed red/gray).  Detector cuts of $|y_j| \le 2.5$ and $p_T^j \ge 25\,\text{GeV}$ have been applied for all distributions.}}
\end{center}
\end{figure}

Similarly to what we observed for the Tevatron, the observables at the LHC are strongly dependent on the jet kinematics. In particular, the jet rapidity distribution of the charge asymmetry and the cross section is determined by the collinear behavior. In Figure~\ref{fig:lhc_ac_sigmas}, we show $A_C^{|y|}$ (left) and $\sigma_s$ (right) as a function of the jet rapidity in the parton frame, $\hat{y}_j$, for the detector cuts $|y_j| \le 2.5$ and $p_T^j \ge 25\,\text{GeV}$. Without further cuts (black curve), the charge asymmetry reaches its maximum for a central jet, $A_C^{|y|}(\hat{y}_j = 0) = -1.5\,\%$. With respect to the total asymmetry $A_C^{|y|} = -0.56\,\%$, the asymmetry is enhanced by almost a factor of three, but remains a small quantity due to the high $gg$ suppression. As was explained in Section~\ref{sec:phasespace}, the charge-symmetric background can be reduced by filtering out highly-boosted $q\bar q$ events through a lower cut on the boost rapidity $y_{t\bar tj}$ defined in Appendix~\ref{app:phasespace}. For instance, by restricting the rapidity to $|y_{t\bar t j}| \ge 0.5$ (dashed red/gray curve), the maximal charge asymmetry increases to $A_C^{|y|}(\hat{y}_j = 0,|y_{t\bar tj}| \ge 0.5) = -2.3\,\%$. The cross section is simultaneously reduced by a factor of two.
%%%%%%%%%%%%%%%%%%%%%%%%%%%%%%%%%%%%%%%%%%%%%%%%%%%%%%%%%%%%%
%%%%%%%%%%%%%%%%%%%%%%%%%%%%%%%%%%%%%%%%%%%%%%%%%%%%%%%%%%%%%

\begin{figure}[!t]
\begin{center}
\includegraphics[height=5cm]{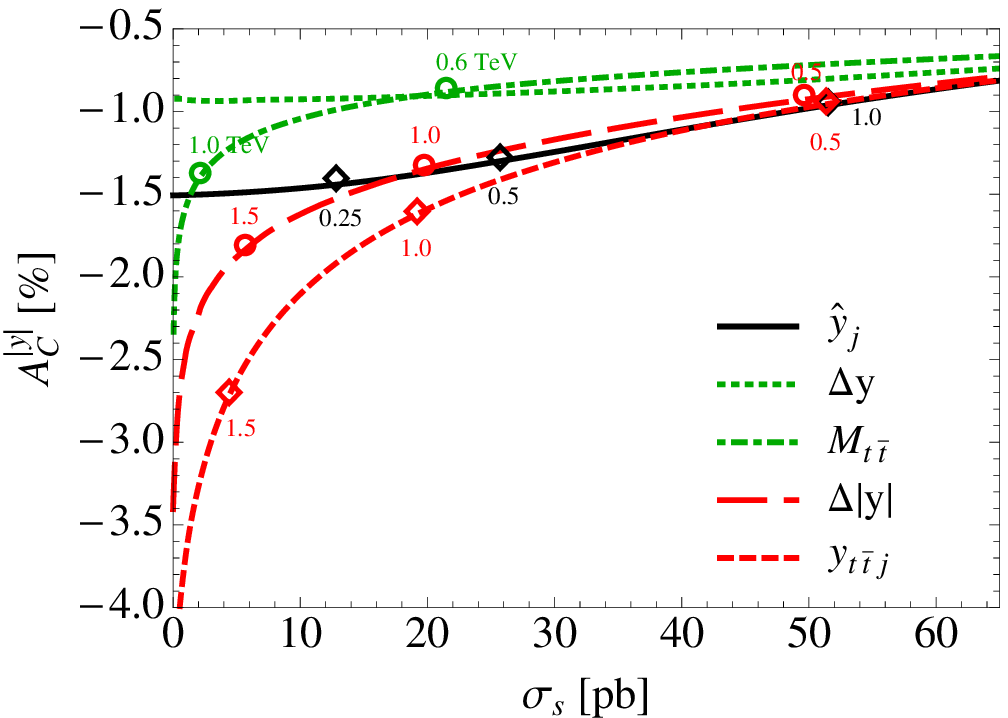}
\hspace*{1.5cm}
\includegraphics[height=5cm]{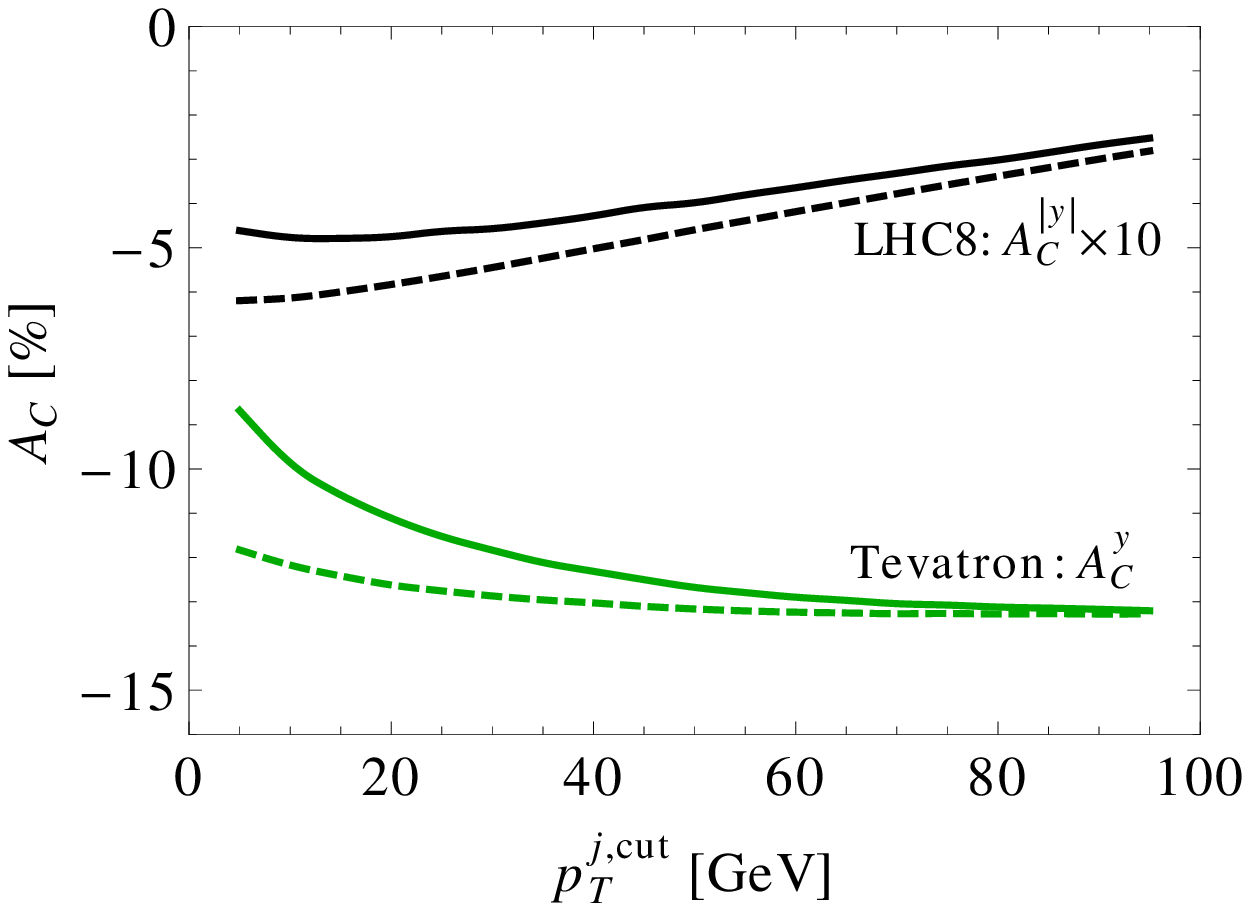}
\end{center}
\vspace*{-1cm}
\begin{center} 
  \parbox{15.5cm}{\caption{\label{fig:lhc_acsigmas_ptj} Left: Integrated charge asymmetry $A_C^{|y|}$ at the LHC8 as a function of the cross section $\sigma_s$ for cuts on kinematic variables $v$. Numbers mark the cuts $v^{\text{min}}$, $v^{\text{max}}$, i.e. $|v| \ge v^{\text{min}}$ ($|v| \le v^{\text{max}}$ for $v=\hat{y}_j$). Right: Charge asymmetry as a function of the minimal jet transverse momentum, $p_T^{j,\text{min}}$, at the Tevatron (green/gray) and the LHC8 (black, rescaled by a factor of $10$), without further cuts (solid) and with the jet rapidity cuts $|y_j| \le 2$ (dashed green/gray) and $|y_j| \le 2.5$ (dashed black).}}
\end{center}
\end{figure}

In addition or alternatively to selecting boosted $q\bar q$ parton contributions, the asymmetry can be enhanced by exploiting the kinematics of the $t\bar t + j$ final state. In the left panel of Figure~\ref{fig:lhc_acsigmas_ptj}, we compare the efficiency of cuts on different variables $v$ to generate a large charge asymmetry $A_C^{|y|}$ for a given cross section $\sigma_s$. We compare cuts on the jet rapidity $\hat{y}_j$, the top-antitop rapidity difference $\Delta y$ and invariant mass $M_{t\bar t}$, which were already relevant for the Tevatron asymmetry. In addition, we introduce cuts on the difference between absolute top and antitop rapidities $\Delta |y|$ and the parton-frame boost rapidity $y_{t\bar t j}$, which are adapted to the experimental setup at the LHC. The numbers in the figure indicate the cuts $v^{\text{min}}$, $v^{\text{max}}$ corresponding to the point $\{\sigma_s, A_C^{|y|}\}(|v| \ge v^{\text{min}})$ for $v = \Delta y$, $M_{t\bar t}$, $\Delta |y|$, $y_{t\bar t j}$ and $\{\sigma_s,A_C^{|y|}\}(|v| \le v^{\text{max}})$ for $v = \hat{y}_j$.
%%%%%%%%%%%%%%%%%%%%%%%%%%%%%%%%%%%%%%%%%%%%%%%%%%%%%%%%%%%%%
%%%%%%%%%%%%%%%%%%%%%%%%%%%%%%%%%%%%%%%%%%%%%%%%%%%%%%%%%%%%%
As for the Tevatron, the charge asymmetry $A_C^{|y|}$ is not very sensitive to a cut on the invariant mass $M_{t\bar t}$ (dot-dashed green/gray curve), unless one enters the regime where the emitted jet is soft (cf. Figure~\ref{fig:infrared_behavior}). For such a strong cut on $M_{t\bar t}$, however, the cross section is very small. A lower cut on $\Delta y$ (dotted green/gray curve), the variable defining the forward-backward asymmetry at the Tevatron, selects $t\bar t$ pairs that are emitted back-to-back and parallel to the beam axis. At the LHC, this cut is not efficient, because the definition of the charge asymmetry in terms of $\Delta |y|$ involves top and antitop quarks that are scattered perpendicular to the beam axis. It is thus more suitable to focus on events where one of the top and antitop quarks is beam-like, while the other quark is central, i.e. on events with a large difference of absolute rapidities $\Delta |y|$. For cross sections $\sigma_s \ge 20 \,\text{pb}$ (corresponding to $\Delta|y|^{\text{min}} \le 1.0$), a lower cut on $\Delta|y|$ (long-dashed red/gray curve) is almost as efficient in increasing $A_C^{|y|}$ as the jet rapidity $\hat{y}_j$ (black curve). For the regime close to the kinematic endpoint, where the cross section is very small, the cut on $\Delta|y|$ is more efficient. The most efficient variable is the boost rapidity $y_{t\bar tj}$ (dashed red/gray curve). For cross sections $\sigma_s \ge 30 \,\text{pb}$, cuts on $y_{t\bar tj}$ and $\hat{y}_j$ are of comparable efficiency. With a very strong cut $y_{t\bar tj}\gtrsim 2$, a maximal asymmetry of $A_C^{|y|} \approx -4 \,\%$ can be achieved. As for the Tevatron, the combination of cuts on several variables is more efficient to increase the asymmetry than a stronger cut on one single variable. Combinations of the most efficient variables $y_{t\bar tj}$ and $\Delta |y|$ with $\hat{y}_j$ are given in Figure~\ref{fig:lhc_ac_sigmas} (dashed and long-dashed red/gray curves).
%%%%%%%%%%%%%%%%%%%%%%%%%%%%%%%%%%%%%%%%%%%%%%%%%%%%%%%%%%%%%%%
\begin{table}
\begin{center}
\begin{tabular}{|c|c|c|c|c|c|c|}
\hline
 &  &  &  & \multicolumn{3}{c}{$|y_{t\bar tj}| \ge 1$} \vline\tabularnewline
\cline{5-7}
 & no cuts & $|\hat{y}_{j}| \le 0.5$ & $|\Delta |y|| \ge 1$ & -- & $|\hat{y}_{j}| \le 0.5$ & $|\hat{y}_{j}| \le 0.5 \le |\Delta|y||$\tabularnewline
\hline
\hline 
$A_{C}^{|y|}\,\,[\%]$ & $-0.56$ & $-1.30$ & $-1.35$ & $-1.62$ & $-2.91$ & $-4.04$\tabularnewline
\hline 
$\sigma_{s}\,\,[\text{pb}]$ & $97.5$ & $25.7$ & $19.7$ & $19.2$ & $6.63$ & $4.00$\tabularnewline
\hline
\end{tabular}
\end{center}
\vspace*{-0.5cm}
\begin{center} 
  \parbox{15.5cm}{\caption{\label{tab:lhc_cuts} Charge asymmetry $A_C^{|y|}$ and cross section $\sigma_s$ at LHC8 for cuts on the partonic jet rapidity $\hat{y}_j$, the rapidity difference $\Delta |y| = |y_t| - |y_{\bar t}|$, and the boost of the parton frame $y_{t\bar t j}$. Detector cuts of $|y_j| \le 2.5$ and $p_T^j \ge 25\,\text{GeV}$ have been applied.}}
\end{center}
\end{table}
%%%%%%%%%%%%%%%%%%%%%%%%%%%%%%%%%%%%%%%%%%%%%%%%%%%%%%%%%%%%%%
In Table~\ref{tab:lhc_cuts}, we display $A_C^{|y|}$ and $\sigma_s$ for various cuts corresponding to the points indicated in Figure~\ref{fig:lhc_acsigmas_ptj}, left. To demonstrate the power of combinations of cuts, in the last two columns we give the results for $|y_{t\bar tj}| \ge 1$ with the simultaneous cuts $|\hat{y}_j| \le 0.5$ and $|\hat{y}_j| \le 0.5$, $|\Delta |y|| \ge 0.5$. For similar cross sections, the resulting charge asymmetries of $A_C^{|y|} = -2.91\,\%$ and $-4.04\,\%$ cannot be achieved by a single cut on either of the variables $|y_{t\bar tj}|$, $|\hat{y}_j|$ or $|\Delta |y||$, as can be read off from Figure~\ref{fig:lhc_acsigmas_ptj}, left. Generically, the charge asymmetry hardly exceeds $-5\,\%$, even for very strong cuts on the cross section. Since NLO corrections cause a positive shift, the absolute value of the charge asymmetry beyond LO is expected to be smaller than at LO. Assuming an experimental efficiency of $5\,\%$ in the lepton + jets channel \footnote{Our assumption is based on the efficiencies derived for the measurement of $\sigma_s^{t\bar t j}$ in \cite{ATLAS:2012ttbj}.} and an integrated luminosity of $20\,\text{fb}^{-1}$, one expects about $N=10^5$ $t\bar t + j$ events to be collected at the LHC in 2012. To give a measure of the observability of a charge asymmetry, we define the statistical significance $S = |A_C^{|y|}|/\delta A_C^{|y|}$, where $\delta A_C^{|y|}\approx 1/\sqrt{N}$. For instance, the combination of cuts $|y_{t\bar tj}| \ge 1$ and $|\hat{y}_j|\le 0.5$ from Table~\ref{tab:lhc_cuts} yields $A_C^{|y|} = -2.9\,\%$ and $\delta A_C^{|y|} = 1.2\,\%$, excluding the zero hypothesis with a significance of $S = 2.4$. By combining the data sets collected by ATLAS and CMS, this statistical significance can be raised to $S = 3.4$. Cuts that enhance the charge asymmetry can be optimized in order to maximize the observability of a signal. Further improvement is possible by investigating the purely leptonic and hadronic decay channels. However, a measurement of the QCD charge asymmetry in $t\bar t + j$ production at the LHC will be difficult and will require an excellent control of the systematical uncertainties. Yet, as we will see in Section~\ref{sec:asymmetry-coloroctets}, the smallness of the QCD charge asymmetry can be turned into a virtue in the search for new particles that can generate large effects on $A_C^{|y|}$.

\subsection{Experimental cuts on jet properties}\label{sec:jetcuts}
We conclude our analysis of the QCD charge asymmetry by discussing its dependence on cuts on the jet properties conditioned by the experimental setup. A cut on the jet rapidity $|y_j|$ takes account of the tracking capacity of the inner detector in the forward region, which is limited to $|y_j| \le 2$ at CDF ($|y_j| \le 2.5$ at D{\O}) and to $|y_j| \le 2.5$ at ATLAS ($|y_j| \le 2.4$ at CMS). For the Tevatron, the effect of varying the cut on $|y_j|$ can be estimated by looking at the dashed black curve in Figure~\ref{fig:tev_ac_sigmas}. A stronger cut discards the region where $\sigma_a$ is small and thereby yields a larger asymmetry $A_C^y$. Analogous considerations apply for variations of the jet rapidity cut at the LHC.

A further cut is required on the transverse momentum of the jet, $p_T^j = E_j\,\sin\theta_j$, in order to avoid misidentification due to the pile-up background in the forward region. The cuts used by the experimental collaborations for top-quark pair production are given by $p_T^j \ge 20\,\text{GeV}$ (CDF and D{\O}), $p_T^j \ge 25\,\text{GeV}$ (ATLAS) and $p_T^j \ge 30\,\text{GeV}$ (CMS). The dependence of the top-quark charge asymmetry on $p_T^j$ can be understood by analyzing its asymptotic behavior at low and high jet transverse momenta. At low $p_T^j$, the charge asymmetry is dominated by the collinear limit $\theta_j\rightarrow 0$ described in (\ref{eq:sigmadiv}), yielding $A_C \sim \ln^{-1}(m_t/p_T^j)$. The charge asymmetry therefore increases with increasing $p_T^j$. High transverse momenta $p_T^j$ correspond to high jet energies $E_j$ where the asymmetry is suppressed (see Figure~\ref{fig:infrared_behavior}, left). In this regime, the charge asymmetry thus decreases with increasing $p_T^j$.

In the right panel of Figure \ref{fig:lhc_acsigmas_ptj}, we show the integrated charge asymmetries $A_C^y$ at the Tevatron (solid green/gray curve) and $A_C^{|y|}$ at the LHC8 (solid black curve) as functions of a lower cut on the jet transverse momentum, $p_T^{j,\text{min}}$. At the Tevatron, little phase space is available for the production of a hard jet. The integrated charge asymmetry is dominated by the region of low $p_T^j$, i.e. by its behavior in the collinear limit. A stronger cut on $p_T^j$ therefore increases $A_C^y$ by several percent with respect to the result based on the detector cut $p_T^{j,\text{min}} = 20\,\text{GeV}$. At the LHC, where the phase space for the jet is larger, the high-$p_T^j$ regime is more important. We observe that already for $p_T^{j,\text{min}}\ge 15\,\text{GeV}$ the asymmetry starts to decrease with increasing $p_T^{j,\text{min}}$. It is thus desirable to aim at keeping the experimental cut on $p_T^{j}$ as low as possible, in order not to suppress the charge asymmetry. In practice, the cut on $p_T^j$ is always accompanied by the above-mentioned cut on the jet rapidity $|y_j|$. We display the effect of the detector cuts $|y_j| \le 2$ for the Tevatron (dashed green/gray curve) and $|y_j| \le 2.5$ for the LHC (dashed black curve), which increase the charge asymmetry in particular in the region of small $p_T^{j,\text{cut}}$. The dependence of the asymmetries on a $p_T^{j}$ cut is thus dampened at the Tevatron, but strengthened at the LHC by the detector cut on $|y_j|$.

\section{Charge asymmetry from massive color-octet bosons}\label{sec:asymmetry-coloroctets}
Massive color-octet vector bosons are expected to leave significant imprints on top-quark pair production at hadron colliders. Such ``massive gluons'' arise, for instance, as witnesses of an enlarged gauge symmetry $SU(3)_L\times SU(3)_R$ at higher energies as in models with chiral color \cite{Frampton:1987dn} or topcolor \cite{Hill:1991at}, or as Kaluza-Klein excitations of the QCD gluon from compactified extra dimensions. Generically, the interactions of massive gluons with SM particles relevant for top-antitop production are determined by the following Lagrangian,
\begin{equation}\label{eq:Gginteractions}
\mathcal{L} = - i g_s\,\bar{q}_i\,\gamma^{\mu} G_{\mu}^a T^a \big[g_V^i + \gamma_5\,g_A^i\big] q_i - g_s\,f_{abc}\,\big[(\partial_{\mu} G_{\nu}^a - \partial_{\nu} G_{\mu}^a)\,G^{b\mu} g^{c\nu} + G^{a\mu} G^{b\nu} (\partial_{\mu} g_{\nu}^c)\big]\,,
\end{equation}
where $g_{\mu}^a$ and $G_{\mu}^a$ denote the QCD gluon and massive gluon fields; $T^a$ and $f_{abc}$ are the generators and structure constants of the $SU(3)_C$ gauge group of strong interactions; and $g_V^i$ and $g_A^i$ describe the vector ($V$) and axial-vector ($A$) couplings of the massive gluon to quarks with flavor $i$ in their weak eigenstates. If the triple gauge boson coupling in (\ref{eq:Gginteractions}) arises from gauge-kinetic terms, its strength is fixed by the QCD coupling constant $g_s$ (see also \cite{Chivukula:2011ng}). Gauge invariance further prohibits the coupling of two gluons to an uneven number of massive gluons.\footnote{This argument holds for models with an enlarged gauge symmetry. In warped extra dimensions, the $ggG$ vertex vanishes due to the wave function orthogonality of the gluon and its Kaluza-Klein excitations \cite{Lillie:2007yh}.} The production of a single color-octet vector boson via gluon-gluon fusion is therefore suppressed with respect to quark-antiquark annihilation.

In this section, we investigate the effects of massive gluons in $t\bar t + j$ production at the LHC. The generation of a charge asymmetry from vector and axial-vector contributions is explained in detail. We discuss strategic cuts that enhance massive gluon effects on $A_C^{|y|}$. We further show that the angular distribution of the jet contains information that can be used to distinguish between vector and axial-vector couplings. In our numerical studies, we employ the same inputs and cuts on the jet properties as for the QCD analysis, which were given in Section~\ref{sec:phasespace}.

%%%%%%%%%%%%%%%%%%%%%%%%%%%%%%%%%%%%%%%%%%%%%%%%%%%%%%%%%%%%%%%%%%%%%%%%%%%%%%%%%%%
\begin{figure}[!t]
\begin{center}
\hspace*{0cm}
\includegraphics[height=2.3cm]{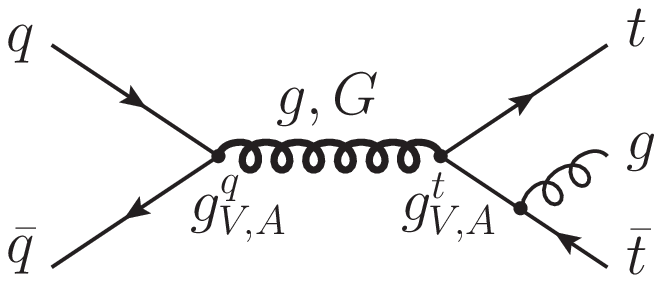}
\hspace*{0cm}
\includegraphics[height=2.3cm]{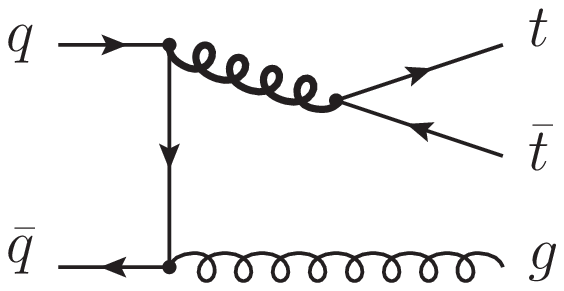}
\hspace*{0cm}
\includegraphics[height=2.3cm]{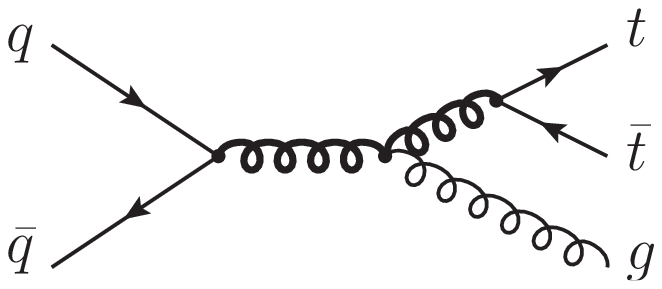}
\vspace*{0.5cm}
\begin{picture}(0,0)(0,0)
\put(-405,-15){\large $(a)$}
\put(-250,-15){\large $(b)$}
\put(-105,-15){\large $(c)$}
\end{picture}
\end{center}
\vspace*{-0.8cm}
\begin{center} 
  \parbox{15.5cm}{\caption{\label{fig:ttj_axi} Leading contributions to the process $q\bar q \rightarrow t\bar t g$ from color-octet vector bosons (bold curled lines) with vector couplings $g_{V}^{q,t}$ and axial-vector couplings $g_{A}^{q,t}$ to quarks. Diagrams for $q g \rightarrow t\bar t q$ are obtained by crossing $\bar{q} \leftrightarrow g$. Diagrams with interchanged $t \leftrightarrow \bar t$ in $(a)$ and $q \leftrightarrow \bar q$ in $(b)$ are not shown.}}
\end{center}
\end{figure}
%%%%%%%%%%%%%%%%%%%%%%%%%%%%%%%%%%%%%%%%%%%%%%%%%%%%%%%%%%%%%%%%%%%%%%%%%%%%%%%%%%%

\subsection{Charge-asymmetric vector and axial-vector contributions}\label{sec:axicontributions}
Massive gluon contributions to the charge-symmetric and -antisymmetric $t\bar t + j$ cross section arise from their interference with the QCD amplitude, $g-G$ ($\sigma_{s,a}^{gG}$), as well as through interference with themselves, $G-G$ ($\sigma_{s,a}^{G}$). They contribute to the top-quark charge asymmetry at the LHC, yielding
\begin{eqnarray}\label{eq:ttbj-observables}
A_C^{|y|,\text{tot}} & = & \frac{\sigma_a^g + \sigma_a^{gG} + \sigma_a^G}{\sigma_s^g + \sigma_s^{gG} + \sigma_s^G} \equiv A_C^{|y|,\text{SM}} + \Delta A_C^{|y|}\,,\qquad \text{with } A_C^{|y|,\text{SM}} = \frac{\sigma_a^g}{\sigma_s^g}.
\end{eqnarray}
The charge asymmetry in $t\bar t + j$ production is generated at LO by the interference of the diagrams shown in Figure~\ref{fig:ttj_axi}. To calculate the contributions to $\sigma_{s,a}^{gG}$ and $\sigma_{s,a}^G$, we use the Feynman rules from \cite{Chivukula:2011ng} and work in the unitary gauge. Notice that the cancellation of unphysical gluon polarization modes requires the inclusion of the diagram $(c)$ with two massive gluons, even though its contribution is suppressed by $\mathcal{O}(M_G^{-2})$ with respect to the diagrams $(a)$ and $(b)$.

Let us discuss the sources of a charge asymmetry from massive gluons in $t\bar t + j$ production in detail. First we focus on the interference of initial- (ISR) and final-state radiation (FSR), i.e., the interference of diagrams $(a)$ and $(b)$ in Figure~\ref{fig:ttj_axi}. Disregarding the color structure, the amplitudes from $V-V$ or $A-A$ (from $V-A$) interference are antisymmetric (symmetric) under $t\leftrightarrow \bar t$. The color structure of ISR-FSR interference can be decomposed in terms of a symmetric factor $d_{abc} = 2\,\text{Tr}[\{T^a,T^b\}T^c]$ and an antisymmetric factor $f_{abc} = -2i\,\text{Tr}[[T^a,T^b]T^c]$. A charge asymmetry can thus be generated in two ways: firstly, from $V-V$ or $A-A$ interference paired with a symmetric color structure $d_{abc}^2$, or secondly by $V-A$ interference and an antisymmetric color structure $f_{abc}^2$. In QCD, the asymmetry is generated by the (antisymmetric) ISR-FSR interference of two vector currents ($g-g$) and is proportional to the symmetric color factor $d_{abc}^2$ \cite{Kuhn:1998kw}. Therefore, the first mechanism is at work. A massive gluon with pure $V$ couplings generates an asymmetry by the same mechanism as in QCD. A massive color octet with $A$ couplings, dubbed axigluon, contributes to the charge asymmetry via both mechanisms. When interfering with itself, the asymmetry is generated via the first mechanism, i.e. by the antisymmetric amplitude. When interfering with the QCD gluon or a massive color octet with $V$ couplings, the second mechanism applies and the antisymmetric color structure is the source of the asymmetry. Besides ISR-FSR interference, a charge asymmetry can be generated from all remaining combinations of diagrams in Figure~\ref{fig:ttj_axi} through the interference of a $V$ and an $A$ current. The additional emitted gluon is just a ``spectator'' of the asymmetry generated by the $V-A$ interference.
%%%%%%%%%%%%%%%%%%%%%%%%%%%%%%%%%%%%%%%%%%%%%%%%%%%%%%%%%%%%%%%%%%%%%%%%%%%%%%%%%%%
\begin{table}[!t]
\begin{center}
\renewcommand{\arraystretch}{1.3}
\begin{tabular}{|l||c|c|c||c|c|c|}
\hline
 & $\sigma_s^g$ & $\sigma_s^{gG}$ & $\sigma_s^G$ & $\sigma_a^g$ & $\sigma_a^{gG}$ & $\sigma_a^G$\tabularnewline
\hline
\hline
$(a)-(a),\,(b)-(b)$ & $1$ & $g_V^q g_V^t$ & $(g_{V,A}^q)^2(g_{V,A}^t)^2$ & $-$ & $g_A^q g_A^t$ & $g_V^q g_V^t\,g_A^q g_A^t$\tabularnewline
$(a)-(b):\,f_{abc}^2$ & $1$ & $g_V^q g_V^t$ & $(g_{V,A}^q)^2(g_{V,A}^t)^2$ & $-$ & $g_A^q g_A^t$ & $g_V^q g_V^t\,g_A^q g_A^t$\tabularnewline
$(a)-(b):\,d_{abc}^2$ & $-$ & $g_A^q g_A^t$ & $g_V^q g_V^t\,g_A^q g_A^t$ & $1$ & $g_V^q g_V^t$ & $(g_{V,A}^q)^2(g_{V,A}^t)^2$\tabularnewline
$(c)-(a),\,(c)-(b)$ & $1$ & $g_V^q g_V^t$ & $(g_{V,A}^q)^2(g_{V,A}^t)^2$ & $-$ & $g_A^q g_A^t$ & $g_V^q g_V^t\,g_A^q g_A^t$\tabularnewline
$(c)-(c)$ & $1$ & $g_V^q g_V^t$ & $(g_{V,A}^q)^2(g_{V,A}^t)^2$ & $-$ & $g_A^q g_A^t$ & $g_V^q g_V^t\,g_A^q g_A^t$\tabularnewline
\hline
\end{tabular}
\end{center}
\begin{center}
  \parbox{15.5cm}{\caption{\label{tab:sigmaa_axi} Contributions to $t\bar t + j$ production from $g-g$ interference ($\sigma_{s,a}^g$), $g-G$ interference ($\sigma_{s,a}^{gG}$), and $G-G$ interference ($\sigma_{s,a}^G$). Contributions labeled $(a)-(b)$ etc. correspond to interfering diagrams in Figure~\ref{fig:ttj_axi} with gluon ($g$) and massive gluon ($G$) insertions. Leading massive gluon effects in inclusive $t\bar t$ production follow the pattern of the first line.}}
\end{center}
\end{table}
%%%%%%%%%%%%%%%%%%%%%%%%%%%%%%%%%%%%%%%%%%%%%%%%%%%%%%%%%%%%%%%%%%%%%%%%%%%%%%%%%%%
The various contributions to the charge-symmetric and -asymmetric cross sections $\sigma_s$ and $\sigma_a$ are summarized in Table~\ref{tab:sigmaa_axi}. Contributions proportional to $g_V^m g_A^n$ with $mn=11,\,13,\,31$ are omitted in the table, but taken into account in our numerical analysis. Their contribution to $\sigma_s$ and $\sigma_a$ is negligibly small. Contributions from the channels $q\bar q\rightarrow t\bar t g$ and $qg \rightarrow t\bar t q$ exhibit the same behavior in terms of quark couplings $g_{V,A}^{q,t}$, albeit with different kinematic coefficients.

Results for the contributions to the charge asymmetry from ISR-FSR interference, labeled $(a)-(b)$ in lines two and three of Table~\ref{tab:sigmaa_axi}, have been previously obtained in \cite{Ferrario:2009ee}. Our results agree with the analytic expressions given in the appendix therein. In our analysis, we calculated the complete set of massive gluon contributions to hadronic $t\bar t + j$ production. The analytic results can be readily obtained from the authors.

\subsection{Benchmark scenarios}\label{sec:benchmarks}
To estimate the effects of massive gluons on the charge asymmetry in $t\bar t + j$ production at the LHC, we define a set of benchmark points given in Table~\ref{tab:benchmarks}. The mass of the color octet is set to $M_G=2\,\text{TeV}$, and its width $\Gamma_G$ is fixed by assuming the decay to SM fermions only (see (47) in \cite{Haisch:2011up}). We distinguish between scenarios with $V$ and $A$ couplings to quarks. The couplings to light quarks $g_{V,A}^q$ are suppressed in order to respect the bounds from dijet resonance searches at the LHC~\cite{ATLAS:dijets,CMS:dijets}. For the scenarios $\va^{xy}$ of massive gluons with both $V$ and $A$ couplings to quarks, the magnitude is chosen such that the combination $(g_V^{q,t})^2 + (g_A^{q,t})^2$ is (almost) the same as for the scenarios $V^x$, $A^x$. Constraints from dijet and $t\bar t$ resonance searches \cite{Aaltonen:2011ts,Abazov:2011gv,ATLAS:2012tx,CMS:ttboost}, which are mostly sensitive to this combination of couplings, are thereby likewise respected. Scenarios with different signs of $g_{V,A}^q\cdot g_{V,A}^t$, marked accordingly by an index $+$ or $-$, probe positive and negative interference with the QCD amplitude, i.e. the sign of $\sigma_{s,a}^{gG}$ in Table~\ref{tab:sigmaa_axi}. In the case of both $V$ and $A$ couplings, we define a third scenario $\va^{+-}$, which probes in addition the sign of the $G-G$ interference terms $\sigma_{s,a}^{G}$. Scenarios with $\va$ structures arise naturally from models with chiral couplings. In particular, a massive gluon that couples only to right-handed (left-handed) quarks yields same-sign (opposite-sign) couplings $g_V$ and $g_A$, as implemented in $\va^{\pm\pm}$ ($\va^{+-}$). All benchmarks are in agreement with the current bounds from charge-symmetric observables at the Tevatron and the LHC \cite{Haisch:2011up}. They further are not restricted by top polarization and spin correlation observables recently measured at the LHC \cite{Krohn:2011tw,ATLAS:toppol,CMS:toppol,ATLAS:2012ao,CMS:spincorr}. Benchmarks with axial-vector couplings, in particular $A^-$ and $\va^{\pm-}$, offer an explanation of the enhanced forward-backward asymmetry measured at the Tevatron \cite{Frampton:2009rk}. Massive gluons lighter than $2\,\text{TeV}$ are strongly constrained by the above-mentioned observables, as well as by measurements of the invariant mass spectrum of $t\bar t$ production at the Tevatron and the LHC \cite{Aaltonen:2009iz,ATLAS:2012hg,CMS:ttbspectrum}. Those constraints can only be relaxed, if the couplings to both light quarks and top quarks, $g_{V,A}^q$ and $g_{V,A}^t$, are strongly suppressed and if the width $\Gamma_G$ is large \cite{Tavares:2011zg}. Light massive gluons will not be considered in this analysis.

%%%%%%%%%%%%%%%%%%%%%%%%%%%%%%%%%%%%%%%%%%%%%%%%%%%%%%%%%%%%%%%%%%%%%%%%%%%%%%%%%%%
\begin{table}[!t]
\begin{center}
\renewcommand{\arraystretch}{1}
\begin{tabular}{|c||c|c|c|c|c|c|c|}
\hline
$M_G=2\,\text{TeV}$ & $V^+$ & $V^-$ & $A^+$ & $A^-$ & $\va^{++}$ & $\va^{--}$ & $\va^{+-}$\tabularnewline
\hline
\hline
$g_V^q$ & $0.5$ & $0.5$ & $0$ & $0$ & $0.35$ & $0.35$ & $0.35$\tabularnewline
$g_V^t$ & $2$ & $-2$ & $0$ & $0$ & $1.5$ & $-1.5$ & $1.5$\tabularnewline
$g_A^q$ & $0$ & $0$ & $0.5$ & $0.5$ & $0.35$ & $0.35$ & $0.35$\tabularnewline
$g_A^t$ & $0$ & $0$ & $2$ & $-2$ & $1.5$ & $-1.5$ & $-1.5$\tabularnewline
\hline
$\Gamma_G/M_G\,[\%]$ & $17.7$ & $17.7$ & $17.3$ & $17.3$ & $19.4$ & $19.4$ & $19.4$\tabularnewline
\hline
\end{tabular}
\end{center}
\begin{center}
 \parbox{15.5cm}{\caption{\label{tab:benchmarks} Benchmark points for massive gluons in $t\bar t + j$ production. Benchmarks are labeled $X^{\text{sgn}}$, where $X=V,A,\va$ stands for vector, axial-vector or both couplings to quarks, and $\text{sgn}=\pm$ refers to the sign of the product $g_{V,A}^q\cdot g_{V,A}^t$. All scenarios respect the constraints from Tevatron and LHC measurements.}}
\end{center}
\end{table}
%%%%%%%%%%%%%%%%%%%%%%%%%%%%%%%%%%%%%%%%%%%%%%%%%%%%%%%%%%%%%%%%%%%%%%%%%%%%%%%%%%%

\subsection{Massive gluon effects close to the resonance}\label{sec:axicuts}
As in QCD, contributions from massive gluons to the LHC asymmetry are suppressed by the large symmetric $gg$ background entering the normalization of $A_C^{|y|}$. Massive gluon effects are further suppressed, because the average partonic CM energy in $t\bar t + j$ production at LHC8 is as low as $\sqrt{\hat{s}}\approx 500\,\text{GeV}$. This is well below the resonance region of $\sqrt{\hat{s}} \approx 2\,\text{TeV} = M_G$, where massive gluons could yield large effects. Effects on the total asymmetry $A_C^{|y|}$ in $t\bar t + j$ production at the LHC are therefore small, if no cuts are applied. In the following, we will discuss how the kinematic cuts defined in Section~\ref{sec:phasespace} can enhance the charge asymmetry from massive gluons.

The suppression due to the $gg$ background, as already described in Section~\ref{sec:phasespace}, is relieved by selecting events with a large boost along the beam line. A lower cut on the total rapidity $y_{t\bar tj}$ projects on the (boosted) $q\bar q$ and $qg$ initial states. Both QCD and massive gluon contributions to the charge-asymmetric cross section $\sigma_a$ are thereby enhanced with respect to the production cross section $\sigma_s$. The enhancement of $A_C^{|y|}$ through a cut on $y_{t\bar tj}$ is largely independent from the massive gluon properties, up to a subleading effect on $\sigma_s$ that enters the normalization and differs for $V$ or $A$ couplings to quarks.

\begin{figure}[!t]
\begin{center}
\includegraphics[height=5cm]{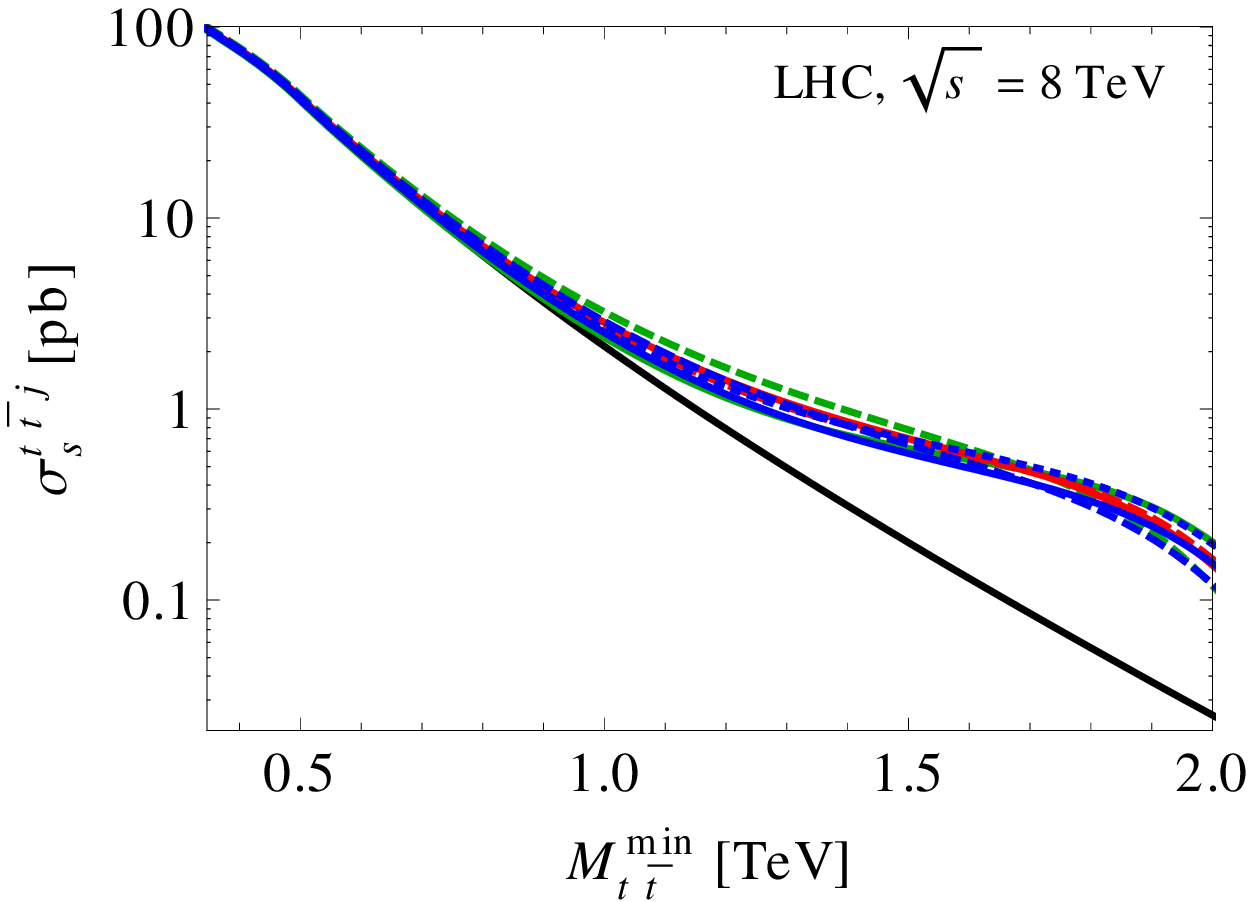}
\hspace*{0.7cm}
\raisebox{2.75cm}{
\begin{tabular}{|c|c|c|}
\hline
$M_{t\bar t}^{\text{min}} = 1\,\text{TeV}$ & $\sigma_s^{\text{tot}}/\sigma_s^g - 1$ & $\Delta A_C^{|y|}$ $[\%]$\tabularnewline
\hline
\hline
$V^+,V^-$ & $0.11,\,0.51$ & $-0.22,\,-2.3$\tabularnewline
$A^+,A^-$ & $0.33,\,0.27$ & $-6.7,\,+4.3$\tabularnewline
$\va^{++}$ & $0.18$ & $+5.4$ \tabularnewline
$\va^{--}$ & $0.35$ & $+8.9$ \tabularnewline
$\va^{+-}$ & $0.23$ & $-6.9$ \tabularnewline
\hline
\end{tabular}}
\end{center}
\vspace*{-1cm}
\begin{center} 
  \parbox{15.5cm}{\caption{\label{fig:csnp_axi8} Left: Massive gluon contributions to the charge-symmetric cross section $\sigma_s^{t\bar t j}$ at LHC8 with a lower cut on the invariant mass, $M_{t\bar t}^{\text{min}}$, for all benchmarks from Table~\ref{tab:benchmarks} (colored as in Figure~\ref{fig:asnp_axi8}). The QCD cross section (black curve) is displayed at LO. Cuts on the jet kinematics $p_T^j \ge 25\,\text{GeV}$ and $|y_j| \le 2.5$ are applied. Right: Massive gluon effects on $\sigma_s^{t\bar t j}$ and $A_C^{|y|, t\bar t j}$ for $M_{t\bar t}^{\text{min}} = 1\,\text{TeV}$ and the benchmarks from Table~\ref{tab:benchmarks}.}}
\end{center}
\end{figure}

\begin{figure}[!t]
\begin{center}
\includegraphics[height=5cm]{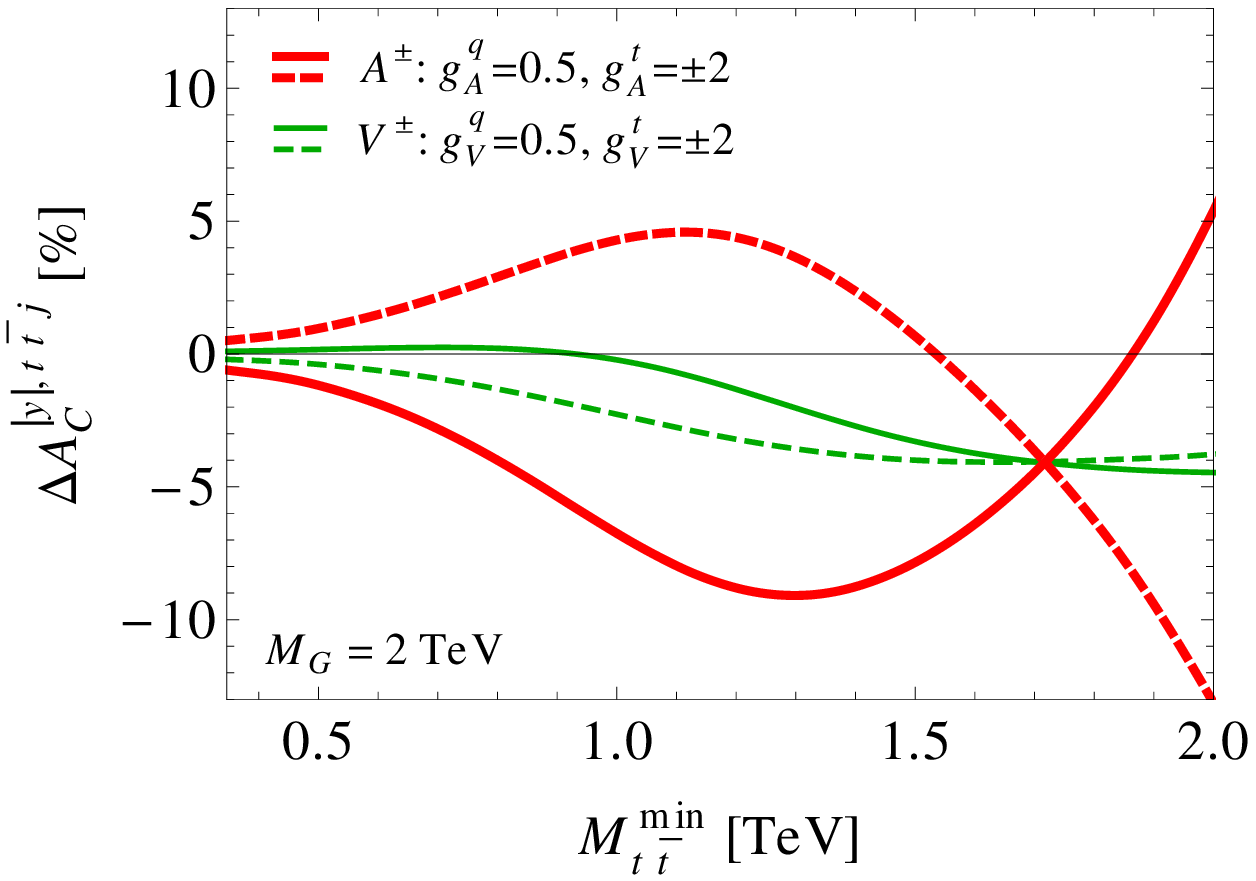}
\hspace*{0.5cm}
\includegraphics[height=5cm]{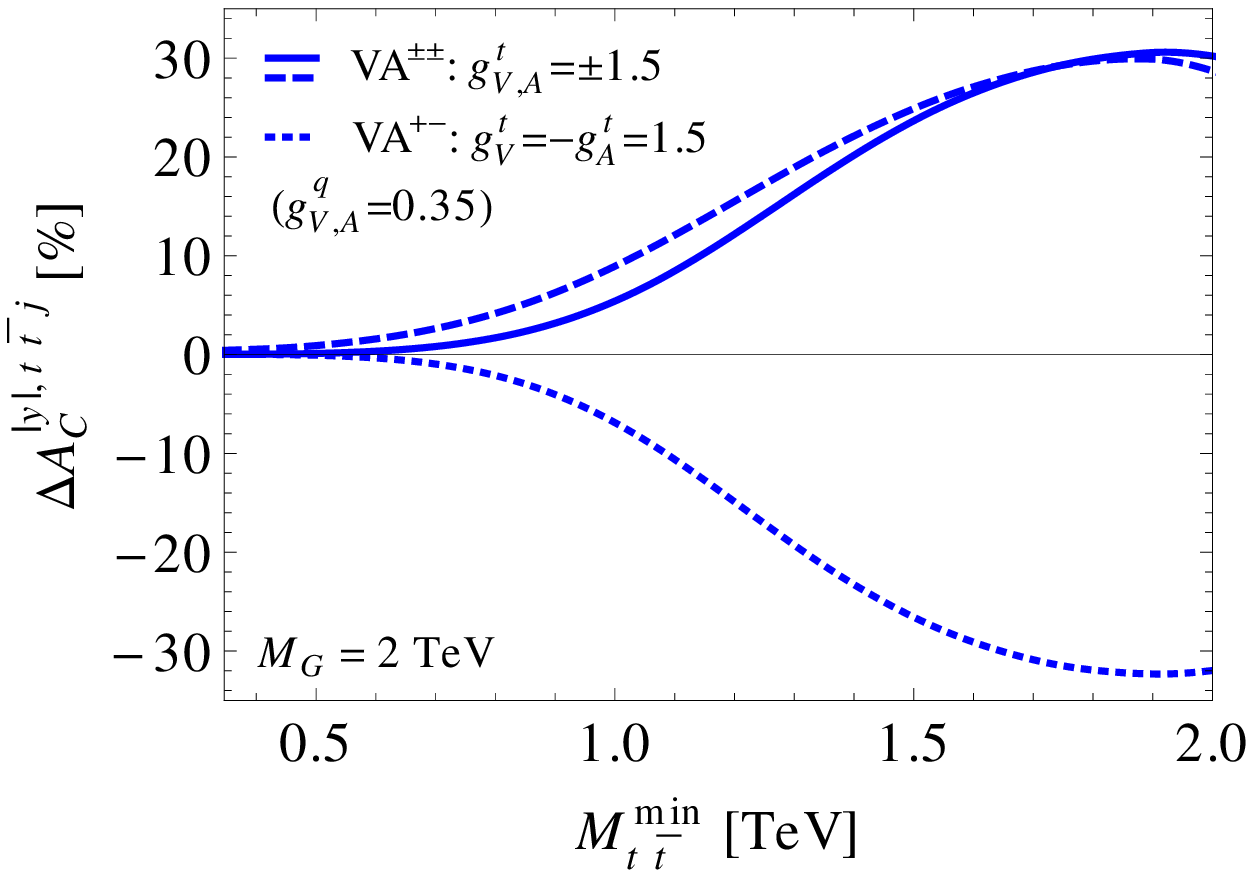}
\end{center}
\vspace*{-1cm}
\begin{center} 
  \parbox{15.5cm}{\caption{\label{fig:asnp_axi8} Massive gluon contributions to the exclusive charge asymmetry $A_C^{|y|,t\bar t j}$ at LHC8 with a lower cut $M_{t\bar t}^{\text{min}}$ for the benchmark scenarios defined in Table~\ref{tab:benchmarks}.}}
\end{center}
\end{figure}

An efficient way to increase massive gluon effects on $A_C^{|y|}$ is to apply a lower cut on the top-antitop invariant mass $M_{t\bar t}$. This enhances the partonic CM energy and brings it closer to the region $M_{t\bar t}\approx M_G$, where massive gluons are produced on shell. The $M_{t\bar t}$ spectrum of the asymmetry is highly model-dependent and sensitive to the interplay of $g-G$ and $G-G$ interference effects. In Figures~\ref{fig:csnp_axi8} and \ref{fig:asnp_axi8}, we show the effects of massive gluons on the symmetric cross section and on the charge asymmetry, $\Delta A_C^{|y|}$ defined in (\ref{eq:ttbj-observables}), at LHC8 as a function of a lower cut on the invariant mass, $M_{t\bar t}^{\text{min}}$. Massive gluon contributions are displayed for the seven benchmark points defined in Table~\ref{tab:benchmarks}. In the symmetric cross section $\sigma_s$ (Figure~\ref{fig:csnp_axi8}), resonance effects of massive gluons show up in the tail of the $M_{t\bar t}$ spectrum. They are largely independent from the respective couplings to quarks. Close to the resonance region, the massive gluon contribution is dominated by $\sigma_s^G$  (cf. Table~\ref{tab:sigmaa_axi}), which in large part does not distinguish between $V$ and $A$ couplings.

The charge asymmetry $A_C^{|y|}$ (Figure~\ref{fig:asnp_axi8}), in turn, is well suited to disentangle massive gluon contributions. For moderate cuts on $M_{t\bar t}$, the effects of pure $V$ or $A$ couplings are driven by the interference with QCD, $\sigma_a^{gG}$. An axigluon with same-sign couplings to light quarks and top quarks ($A^+$, thick solid red/gray curve in Fig. 9, left) interferes constructively with QCD, yielding a negative asymmetry. An axigluon with opposite-sign couplings ($A^-$, thick dashed red/gray curve in Fig. 9, left) interferes destructively and generates a positive asymmetry. Vector contributions (green/gray curves in Fig. 9, left) to $\sigma_a^{gG}$ are much smaller in magnitude and of opposite sign with respect to axial-vector ones. In scenarios with both $V$ and $A$ couplings (blue/black curves in Fig. 9, right), the interference term $\sigma_a^{gG}$ can be of either sign, depending on the dominance and flavor dependence of $V$ or $A$ contributions. Those $\va$ scenarios feature additional contributions to $\sigma_a^G$, which are large both in the resonance region and closer to the top-pair production threshold. They are typically positive, but can be negative if $V$ and $A$ couplings to light quarks \textit{or} top quarks are of opposite sign, as is realized in the benchmark $\va^{+-}$ (dotted blue/black curve in Fig. 9, right). In all scenarios, the interference term changes its sign at $M_{t\bar t}^{\text{min}} = 1.72\,\text{TeV}$, such that the massive gluon contribution is completely determined by $\sigma_a^G$. The dominance of $\va$ contributions to $\sigma_a^G$ in this region is apparent when comparing with pure $V$ or $A$ contributions. In the latter case, one finds $\Delta A_C^{|y|} = -4\%$ for the scenarios $V^{\pm}$, $A^{\pm}$, which is exceeded by the much larger contribution of $\Delta A_C^{|y|} = + 29\,\%$ ($- 31\,\%$) for the scenarios $\va^{\pm\pm}$ ($\va^{+-}$). In the table in Figure~\ref{fig:csnp_axi8}, we give results for massive gluon contributions to the cross section $\sigma_s^{\text{tot}} = \sigma_s^g + \sigma_s^{gG} + \sigma_s^G$ and the charge asymmetry $A_C^{|y|}$ for $M_{t\bar t}^{\text{min}} = 1\,\text{TeV}$. They illustrate once more that the cross section can be significantly enhanced in all massive gluon scenarios at high $M_{t\bar t}$, and that contributions to the charge asymmetry are sizeable and strongly model-dependent.

The total LHC asymmetry $A_C^{|y|,\text{tot}}$ is obtained by adding the QCD asymmetry $A_C^{|y|,\text{SM}}$ to the massive gluon effects $\Delta A_C^{|y|}$ shown in Figure~\ref{fig:asnp_axi8}. As we saw in Section~\ref{sec:abclhc}, the QCD asymmetry at LO is small and negative and increases slightly at high $M_{t\bar t}$. For $M_{t\bar t}^{\text{min}} \le 1\,\text{TeV}$, $A_C^{|y|,\text{SM}}$ stays below $-1.5\,\%$ and does not exceed $A_C^{|y|,\text{SM}} = -2.5\,\%$ for even stronger cuts on $M_{t\bar t}$, as can be observed from Figure~\ref{fig:lhc_acsigmas_ptj}, left. Furthermore, NLO corrections decrease the absolute value of the QCD asymmetry. In the region of high $M_{t\bar t}$, the total asymmetry is thus dominated by massive gluon effects, $A_C^{|y|,\text{tot}} \approx \Delta A_C^{|y|}$, and can be read off to a good approximation from Figure~\ref{fig:asnp_axi8}. For $M_{t\bar t}^{\text{min}}=1\,\text{TeV}$, an efficiency of $5\,\%$ and $20\,\text{fb}^{-1}$ luminosity at LHC8, the statistical error on the asymmetry is expected to be about $\delta A_C^{|y|}(M_{t\bar t}^{\text{min}} = 1\,\text{TeV}) = 2\,\%$, assuming that the QCD cross section $\sigma_s(M_{t\bar t}^{\text{min}}=1\,\text{TeV}) = 2\,\text{pb}$ is not significantly changed by massive gluon contributions. Then, effects from massive color octets of at least $\Delta A_C^{|y|} = \mathcal{O}(\pm 6\,\%)$ are observable with a statistical significance of three standard deviations. By comparing with Figure~\ref{fig:csnp_axi8}, right, we conclude that the benchmark models with axial-vector or chiral couplings to quarks ($A, \va$) should be observable at the LHC with 2012 data. As mentioned before, the combination of different top-quark decay channels and of the results from both ATLAS and CMS would allow to detect even smaller effects of new resonances.

%%%%%%%%%%%%%%%%%%%%%%%%%%%%%%%%%%%%%%%%%%%%%%%%%%%%%%%%%%%%%%%%%%%%%%%%%%%%%%%%%
%%%%%%%%%%%%%%%%%%%%%%%%%%%%%%%%%%%%%%%%%%%%%%%%%%%%%%%%%%%%%%%%%%%%%%%%%%%%%%%%%%%
%%%%%%%%%%%%%%%%%%%%%%%%%%%%%%%%%%%%%%%%%%%%%%%%%%%%%%%%%%%%%%%%%%%%%%%%%%%%%%%%%%%
%%%%%%%%%%%%%%%%%%%%%%%%%%%%%%%%%%%%%%%%%%%%%%%%%%%%%%%%%%%%%%%%%%%%%%%%%%%%%%%%%%%
%%%%%%%%%%%%%%%%%%%%%%%%%%%%%%%%%%%%%%%%%%%%%%%%%%%%%%%%%%%%%%%%%%%%%%%%%%%%%%%%%
\begin{figure}[!t]
\begin{center}
\includegraphics[height=5cm]{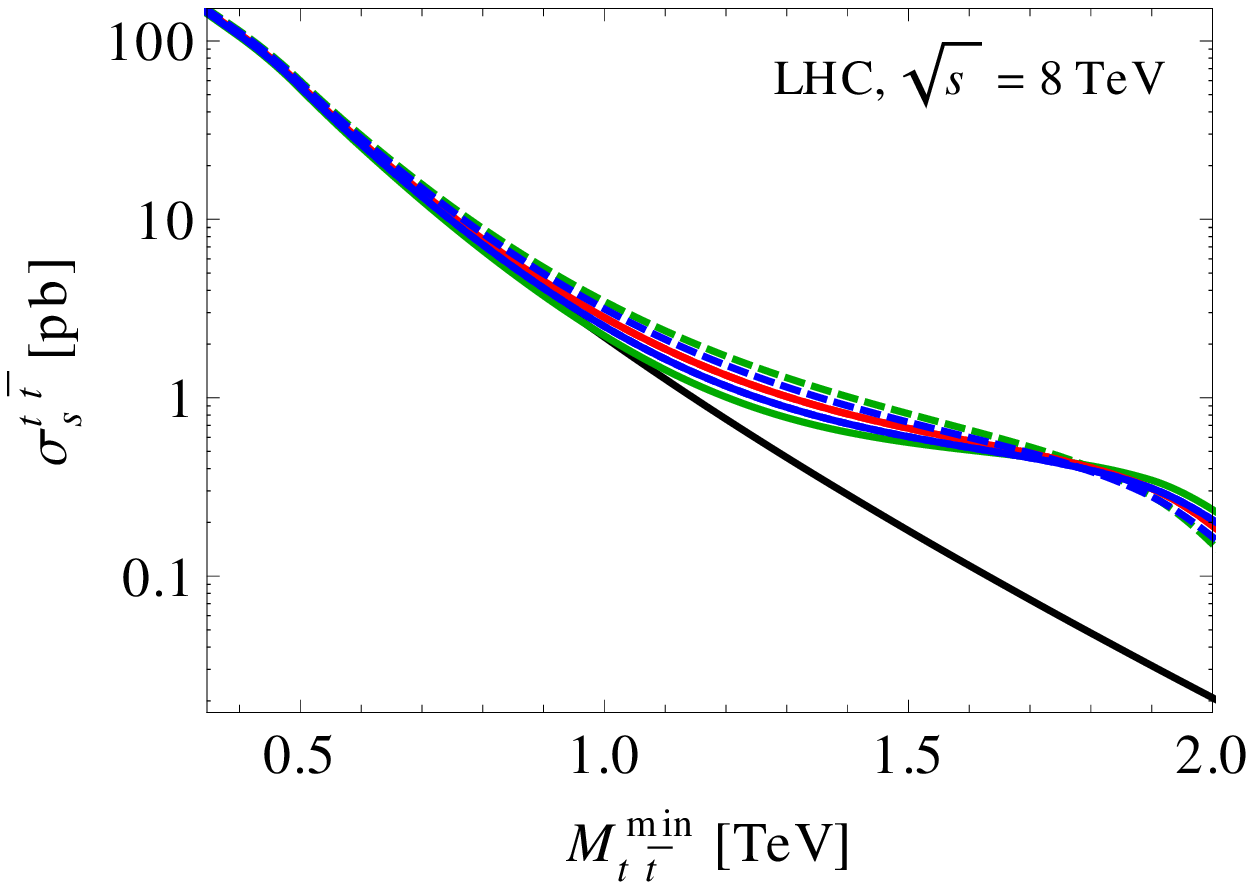}
\hspace*{0.5cm}
\includegraphics[height=5cm]{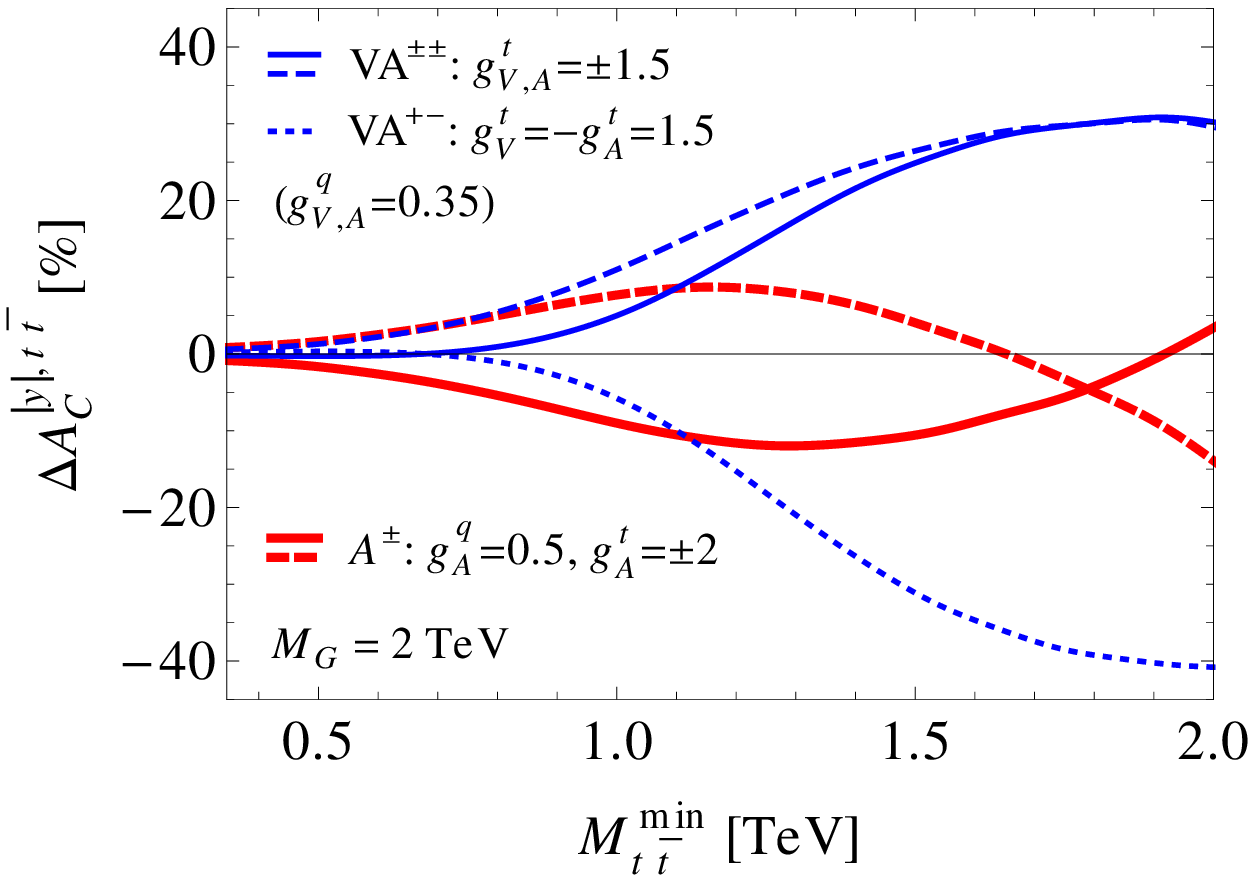}
\end{center}
\vspace*{-1cm}
\begin{center} 
  \parbox{15.5cm}{\caption{\label{fig:asnp_axi8_inc} Massive gluon contributions to the cross section $\sigma_s^{t\bar t}$ (left) and to the charge asymmetry $A_C^{|y|,t\bar t}$ (right) in inclusive $t\bar t$ production at LHC8 for the benchmarks defined in Table~\ref{tab:benchmarks}.}}
\end{center}
\end{figure}
%%%%%%%%%%%%%%%%%%%%%%%%%%%%%%%%%%%%%%%%%%%%%%%%%%%%%%%%%%%%%%%%%%%%%%%%%%%%%%%%%%%

Since the $M_{t\bar t}$ spectrum of the asymmetry nicely resolves the various contributions, we take the opportunity to compare massive gluon effects on the charge asymmetries in exclusive $t\bar t + j$ production, $A_C^{|y|,t\bar t j}$ (Figure~\ref{fig:asnp_axi8}), and in inclusive $t\bar t$ production, $A_C^{|y|,t\bar t}$ (Figure~\ref{fig:asnp_axi8_inc}, right panel). The inclusive asymmetry $A_{C}^{|y|,t\bar t}$ is generated at tree level by the interference of an axigluon with the QCD gluon; all leading effects follow the pattern in the first line of Table~\ref{tab:sigmaa_axi}. Charge-asymmetric contributions from massive gluons with $V$ couplings are absent in inclusive $t\bar t$ production at LO and can thus be probed only in the exclusive asymmetry $A_C^{|y|,t\bar t j}$. Axial-vector contributions to $A_C^{|y|,t\bar t}$ and $A_C^{|y|,t\bar tj}$ are of similar size, but they arise from different sources. In inclusive $t\bar t$ production, the LO asymmetry is generated solely by $V-A$ interference. In $t\bar t + j$ production, the same mechanism is at work for ISR-ISR and FSR-FSR interference terms (Table~\ref{tab:sigmaa_axi}, line 1), which are NLO corrections to the inclusive asymmetry. The dominant contribution, however, stems from ISR-FSR interference (Table~\ref{tab:sigmaa_axi}, line 2), where the asymmetry is generated by the antisymmetric color structure. Both contributions interfere destructively, which reduces the total effect on $A_C^{|y|,t\bar tj}$. In the search for massive color octets, the charge asymmetry in $t\bar t + j$ production has thus more to offer than NLO corrections to the inclusive asymmetry. It additionally probes vector couplings at tree level and effects from diagrammatically different sources.

\subsection{Massive gluon couplings from the jet angular distribution}\label{sec:axicouplings}
\begin{figure}[!t]
\begin{center}
\includegraphics[height=5cm]{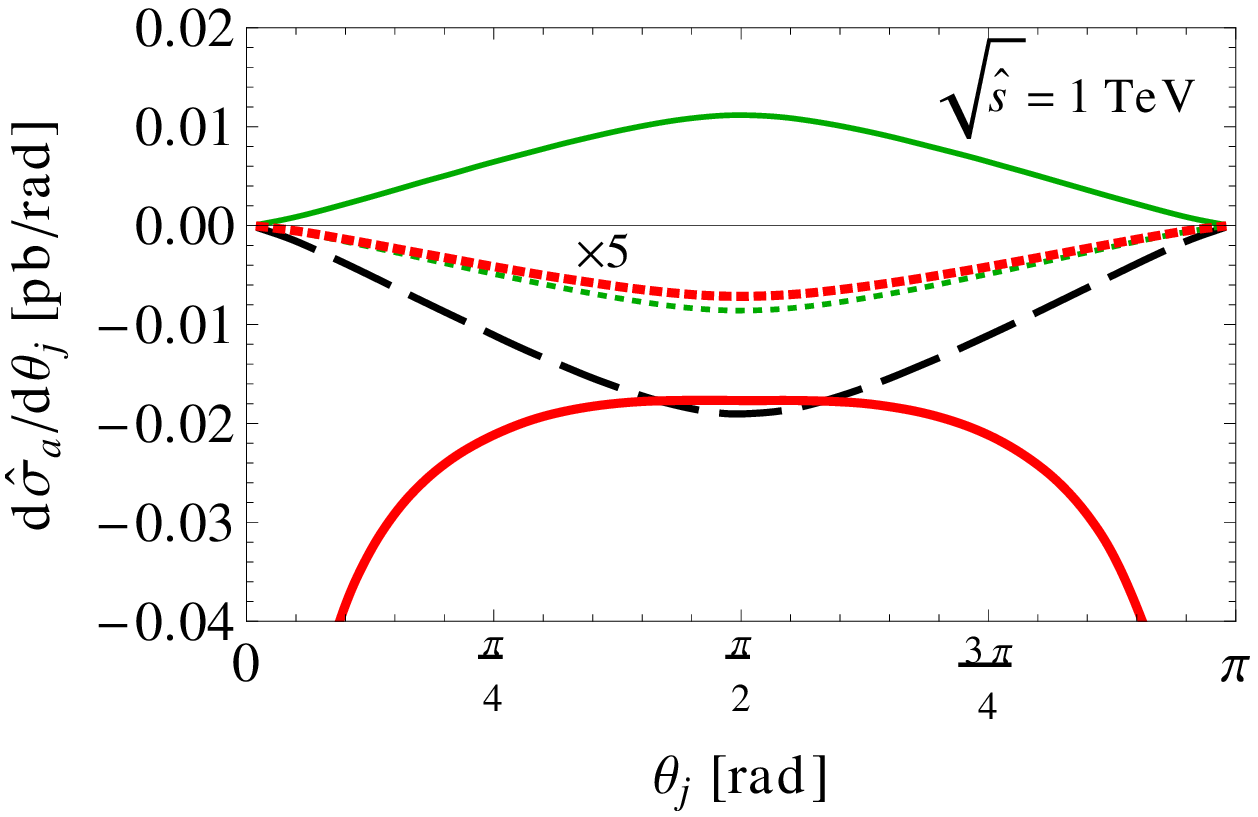}
\hspace*{0.5cm}
\includegraphics[height=5cm]{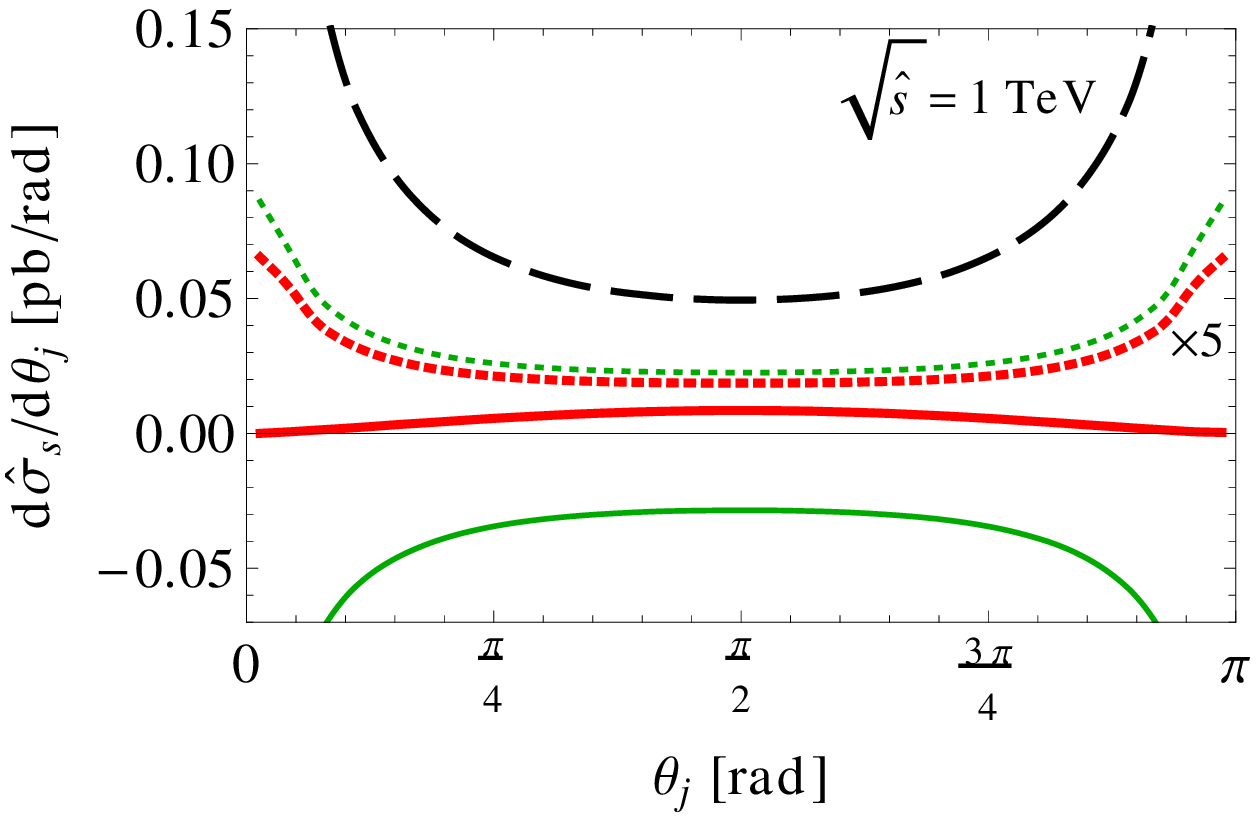}
\end{center}
\vspace*{-1cm}
\begin{center} 
  \parbox{15.5cm}{\caption{\label{fig:jet_dist_axi} Jet angular distribution of the charge-asymmetric (left) and -symmetric (right) partonic $q\bar q\rightarrow t\bar t g$ cross sections for $\sqrt{\hat{s}} = 1\,\text{TeV}$ and $p_T^j \ge 2\,\text{GeV}$. Shown are contributions from QCD (long-dashed black), as well as from $g-G$ (solid) and $G-G$ (dotted, rescaled by a factor $5$) interference for the massive gluon benchmarks $V^+$ (green/gray) and $A^+$ (thick red/gray) defined in Table~\ref{tab:benchmarks}.}}
\end{center}
\end{figure}

The angular distribution of the jet in $t\bar t + j$ production provides complementary information on the massive gluon couplings to the invariant mass spectrum. Like in QCD, the shape of the jet distribution for massive gluon contributions is driven by the behavior of the cross sections in the collinear regime. Figure~\ref{fig:jet_dist_axi} shows the jet angular distributions $\text{d}\hat{\sigma}_a/\text{d}\theta_j$ (left) and $\text{d}\hat{\sigma}_s/\text{d}\theta_j$ (right) defined in (\ref{eq:dsigmadtheta}) for the partonic process $q\bar q\rightarrow t\bar t g$ at a CM energy of $\sqrt{\hat{s}}=1\,\text{TeV}$. For $V-V$ interference (solid and dotted green/gray curves) and $A-A$ interference (thick dotted red/gray curve), the collinear behavior follows that of QCD (long-dashed black curve). The symmetric cross section $\sigma_s$ exhibits a collinear divergence, whereas the asymmetric cross section $\sigma_a$ smoothly vanishes in the collinear limit. For $V-A$ interference (thick solid red/gray curve), the collinear behavior is reversed with respect to QCD: $\sigma_s$ vanishes in the collinear limit and $\sigma_a$ is divergent. In short, only those amplitudes for which the jet determines the properties under charge conjugation (disregarding the color structure) are finite in the collinear limit. From a diagrammatic point of view, only ISR-FSR interference is collinear finite and generates a central jet. All other contributions are collinear divergent, so that the jet is preferentially emitted in the beam direction.
%%%%%%%%%%%%%%%%%%%%%%%%%%%%%%%%%%%%%%%%%%%%%%%%%%%%%%%%%%%%%%%%%%%%%%%%%%%%%%%%%%%
%%%%%%%%%%%%%%%%%%%%%%%%%%%%%%%%%%%%%%%%%%%%%%%%%%%%%%%%%%%%%%%%%%%%%%%%%%%%%%%%%%%
%%%%%%%%%%%%%%%%%%%%%%%%%%%%%%%%%%%%%%%%%%%%%%%%%%%%%%%%%%%%%%%%%%%%%%%%%%%%%%%%%%%
\begin{figure}[!h]
\begin{center}
\includegraphics[height=5cm]{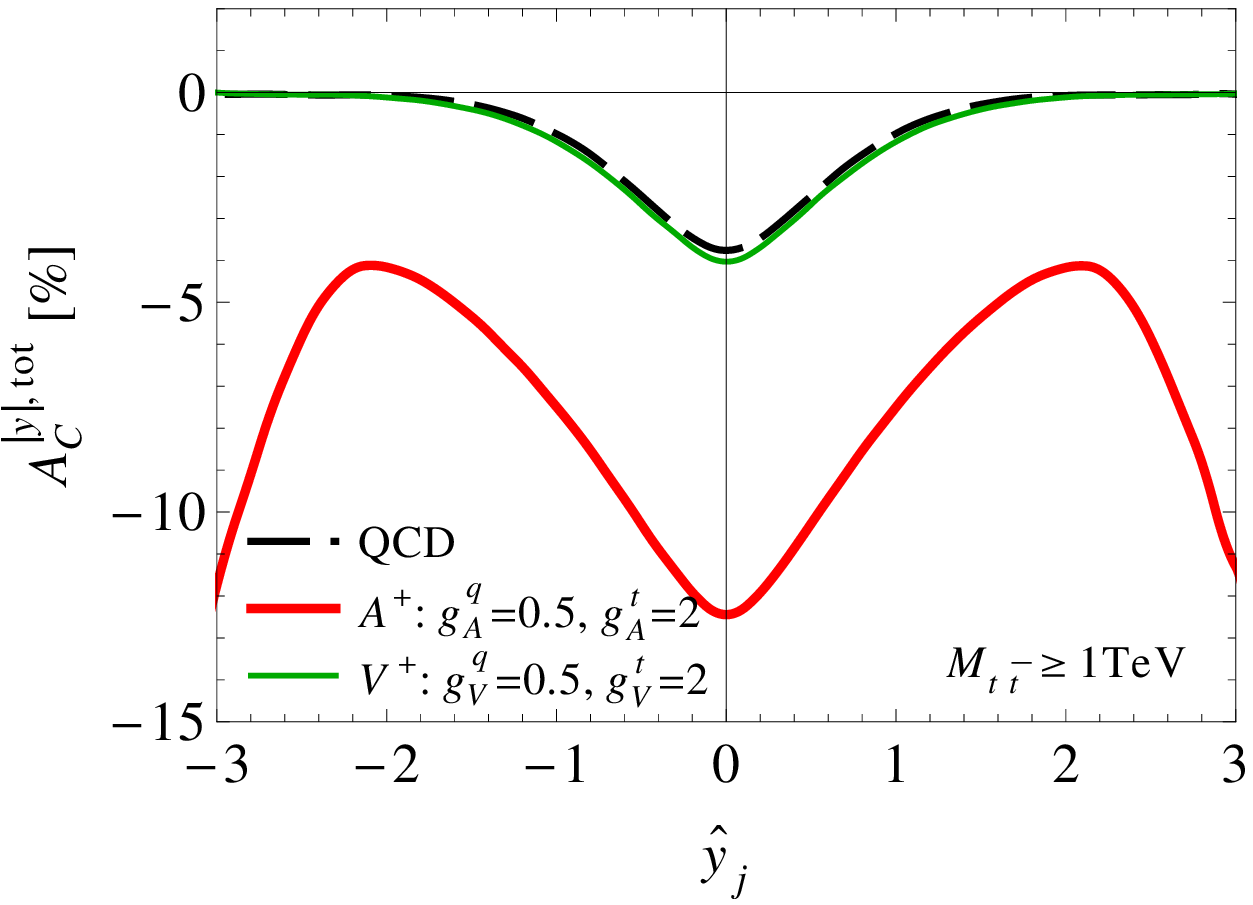}
\hspace*{0.5cm}
\includegraphics[height=5cm]{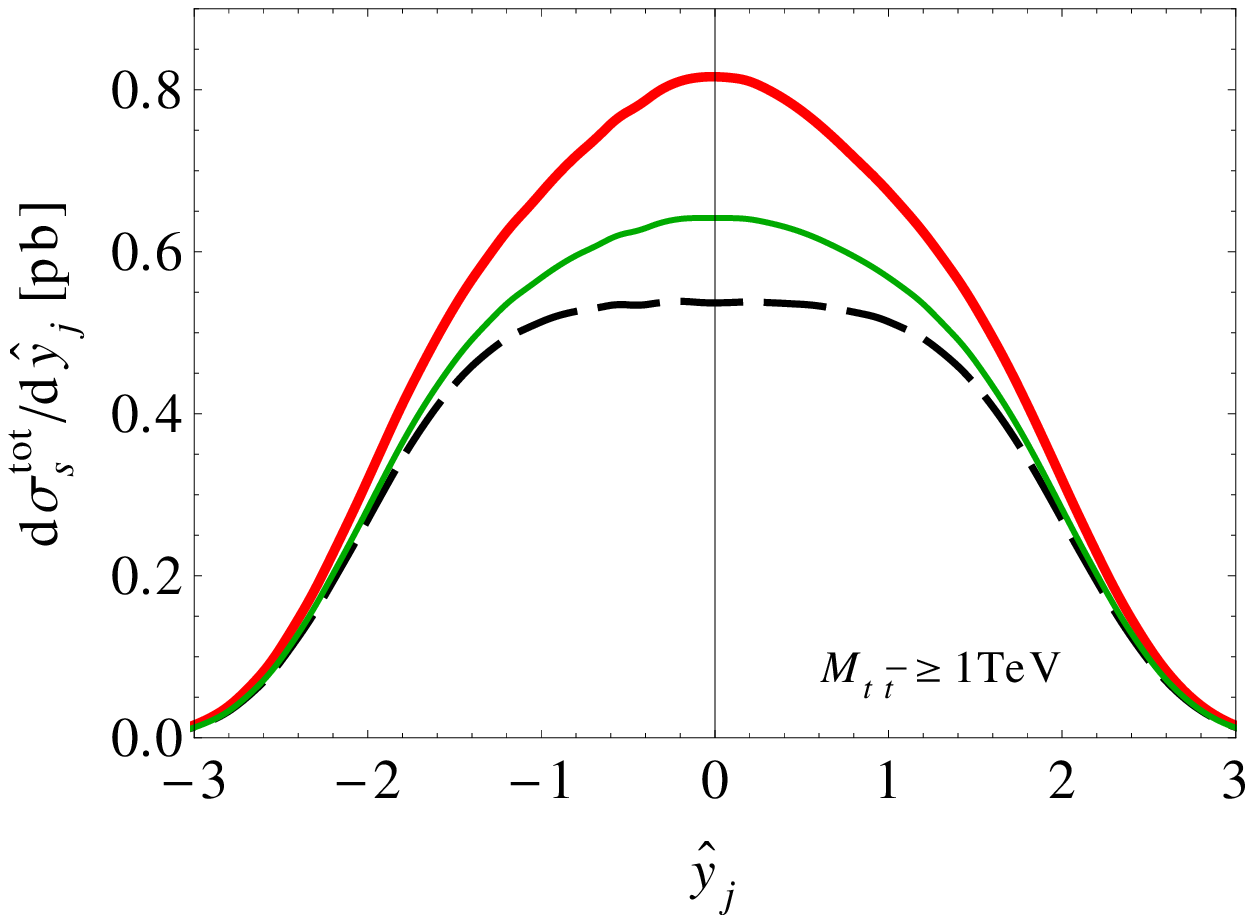}
\end{center}
\vspace*{-1cm}
\begin{center} 
  \parbox{15.5cm}{\caption{\label{fig:jet_dist_asymmetry} Charge asymmetry $A_C^{|y|,\text{tot}}$ (left) and differential cross section $\text{d}\sigma_s^{\text{tot}}/\text{d}\hat{y}_j$ (right) at LHC8 as a function of the rapidity $\hat{y}_j$ for $M_{t\bar t} \ge 1\,\text{TeV}$. Shown is the sum of QCD and massive gluon contributions for the scenarios $V^+$ (green/gray) and $A^+$ (thick red/gray).}}
\end{center}
\end{figure}
%%%%%%%%%%%%%%%%%%%%%%%%%%%%%%%%%%%%%%%%%%%%%%%%%%%%%%%%%%%%%%%%%%%%%%%%%%%%%%%%%%%

The jet angular distribution may serve as a tool to distinguish $V$ from $A$ contributions to the charge asymmetry by exploring their differing collinear behavior. In Figure~\ref{fig:jet_dist_asymmetry}, the LHC asymmetry $A_C^{|y|,\text{tot}}$ (left), defined in (\ref{eq:ttbj-observables}), and the cross section $\sigma_s^{\text{tot}}$ (right) are shown as a function of the rapidity $\hat{y}_j$ for a lower cut on the invariant mass, $M_{t\bar t} \ge 1\,\text{TeV}$. For massive gluons with vector couplings (green/gray curve), just as in QCD (long-dashed black curve), the asymmetry reaches its maximum in the central jet region. The vector contribution slightly increases the maximal asymmetry from $A_C^{|y|}(\hat{y}_j = 0) = -3.7\,\%$ in QCD to $A_C^{|y|}(\hat{y}_j = 0) = -4.0\,\%$. For axial-vector contributions (thick red/gray curve), the shape of the jet distribution is driven by the interplay of QCD and axigluon terms. In the central region, the asymmetry exhibits a maximum of $A_C^{|y|}(\hat{y}_j = 0) = -12.4\,\%$, because $\sigma_a^{\text{tot}}$ is large and negative, while $\sigma_s^{\text{tot}}$ is in its minimum and increases for more forward-emitted jets (cf. Figure~\ref{fig:jet_dist_axi}). A second maximum is obtained in the limit of large $|\hat{y}_j|$, where the collinear enhancement of the axigluon-QCD interference term $\sigma_a^{gG}$ dominates the asymmetry. The appearance of a large charge asymmetry in the forward-jet region is thus a clear signal of axial-vector contributions in $t\bar t + j$ production. Again, a reliable numerical prediction of the observables in this region requires a dedicated study of resummation effects in the collinear limit.

\section{Conclusions}\label{sec:conclusions}
In this work, we have investigated the charge asymmetry in $t\bar t + j$ production for the Tevatron and LHC observables. Jet kinematics of the total cross section and of the charge asymmetry have been explored in detail in terms of the jet energy $E_j$ and the scattering angle $\theta_j$. The jet angular distribution is driven by the behavior of the cross sections $\sigma_s$ and $\sigma_a$ in the collinear limit. In QCD, the charge-asymmetric cross section $\sigma_a$ does not exhibit a logarithmic enhancement in the limit $\theta_j\rightarrow 0$ and reaches its maximum if the jet is emitted perpendicular to the beam axis. This is also the region where the symmetric cross section $\sigma_s$, which normalizes the charge asymmetry and does exhibit a collinear divergence, is less affected by resummation effects. The utility of a selection of central jet emission is thus twofold: It enhances the charge asymmetry $A_C$ and simultaneously yields a collinear-safe observable. The dependence of the charge asymmetry on the jet energy is mild, apart from the region of large $E_j$, where $A_C$ is suppressed.
% Jet kinematics are also the main reason why the charge asymmetry of the partonic process $qg\rightarrow t\bar t q$ is strongly suppressed with respect to $q\bar q\rightarrow t\bar t g$. In the $qg$ channel, the jet is boosted in the direction of the incident quark and carries large parts of the quark momentum. A charge asymmetry, however, is generated mainly from soft jets with small energies $E_j$. At the LHC, the $qg$ channel thus acts rather like a background for the total charge asymmetry in $t\bar t + j$ production, because its contribution to the total cross section is sizeable. At the Tevatron, the $qg$ contribution is less important, because the cross section is dominated by the process $q\bar q\rightarrow t\bar t g$.

These considerations for the jet kinematics can be applied to the charge asymmetries both at the Tevatron and at the LHC. At the Tevatron, the charge asymmetry in QCD at LO is sizeable, $A_C^y = -12.6\,\%$, but reduced to $A_C^y\approx -4\,\%$ by NLO corrections. We found that in the central jet region, the asymmetry is as large as $A_C^y(\hat{y}_j = 0) = -27\,\%$. An upper cut on the jet rapidity in the parton frame, $\hat{y}_j = y_j - y_{t\bar t j}$, is thus suited to enhance the integrated asymmetry. Apart from jet kinematics, the charge asymmetry is sensitive to the top-antitop rapidity difference $\Delta y$, which defines the observable $A_C^y$, and to a lesser extent to the invariant mass $M_{t\bar t}$. Lower cuts on those variables can further enhance the asymmetry, at the cost of strongly reducing the cross section $\sigma_s$. Since the phase space of hard jet emission in $t\bar t$ production at the Tevatron is small, the cross section for $t\bar t + j$ production amounts only to $\sigma_s^{\text{LO}} = 1.42\,\text{pb}$. From an experimental point of view, the application of cuts that reduce the cross section is therefore not welcome, and the measurement of the total charge asymmetry with the full data set of $10\,\text{fb}^{-1}$ luminosity is difficult.

At the LHC, the phase space in $t\bar t$ production and thereby cross section for hard jet emission is much larger due to the high CM energy. At $\sqrt{s} = 8\,\text{TeV}$, the cross section is sizeable, $\sigma_s^{\text{LO}} = 97.5\,\text{pb}$, but the total charge asymmetry $A_C^{|y|} = -0.56\,\%$ is tiny due to the large charge-symmetric $gg$ background. This background can be suppressed by selecting $t\bar t + j$ events with a large boost $\beta$ of the partonic CM frame, which predominantly stem from $q\bar q$ or $qg$ initial states. A lower cut on the correlated lab-frame rapidity $y_{t\bar t j}$, for instance $|y_{t\bar t j}| \ge 1$, enhances the charge asymmetry by a factor of three to $A_C^{|y|} = -1.62\,\%$ and simultaneously reduces the cross section to $\sigma_s = 19.2\,\text{pb}$. The charge asymmetry exhibits a similar sensitivity to jet kinematics for moderate cuts. The maximal asymmetry is reached in the central jet region, $A_C^{|y|}(\hat{y}_j = 0) = -1.5\,\%$. A third possibility to increase the charge asymmetry is via a lower cut on the difference of absolute top and antitop rapidities $\Delta|y|$, which defines the LHC observable $A_C^{|y|}$. Since the cross section of $t\bar t + j$ production at the LHC is sizeable and the data sample for LHC8 is expected to reach about $20\,\text{fb}^{-1}$ in 2012, the application of cuts is possible and indispensable in order to obtain a measurable charge asymmetry. We showed that by combining cuts on three suitable variables, for $|y_{t\bar t j}| \ge 1$, $|\hat{y}_j| \le 0.5$ and $|\Delta |y|| \ge 0.5$ the asymmetry can be enhanced to $A_C^{|y|} = -4.0\,\%$ with $\sigma_s = 4.0\,\text{pb}$. An asymmetry of that size and for a comparable cross section cannot be obtained by applying a stronger cut on only one of the variables. However, even for stronger combined cuts the asymmetry at LHC8 hardly exceeds $-5\,\%$, making its measurement a challenging task.

Massive color-octet bosons, in turn, can generate large effects on the charge asymmetry at the LHC. Both vector and axial-vector couplings contribute to $A_C^{|y|}$ at tree level. An asymmetry from massive gluons with vector couplings to quarks arises just as in QCD from the antisymmetric amplitude for the ISR-FSR interference of two vector currents. With axial-vector couplings, an asymmetry from axigluon-gluon interference can be generated either by the antisymmetric amplitude for ISR-ISR or FSR-FSR interference or by the antisymmetric color structure in the case of ISR-FSR interference. The different contributions from vector and axial-vector couplings can be disentangled after applying a lower cut on the invariant mass, which projects on the resonance region $M_{t\bar t} \approx M_G$, where massive gluon effects are maximal. Contributions from axigluons are generically larger than from massive gluons with vector couplings and depend on the respective sign of the couplings to light quarks and top quarks. For $M_{t\bar t}^{\text{min}} = 1\,\text{TeV}$, they amount to $\Delta A_C^{|y|} = -6.7\,\%$, $+4.3\,\%$, $-0.22\,\%$, $-2.3\,\%$ for the respective benchmark scenarios $A^+$, $A^-$, $V^+$ and $V^-$ defined in Table~\ref{tab:benchmarks}. The effects are mostly driven by the QCD-massive gluon interference term $\sigma_a^{gG}$. At high $M_{t\bar t}$, the massive-gluon term $\sigma_a^G$ yields large effects, if both vector and axial-vector couplings (i.e. chiral couplings) are present. Contributions to the charge asymmetry for $M_{t\bar t}^{\text{min}} = 1.72\,\text{TeV}$, where $\sigma_a^{gG}$ vanishes, can be as large as $\Delta A_C^{|y|} = + 29\,\%$ or $- 31\,\%$ for the benchmarks $\va^{\pm\pm}$ or $\va^{+-}$. Thus, if the charge asymmetry is enhanced by an axial-vector or a chiral massive gluon contribution, it should be observable at the LHC. The jet angular distribution can be used to obtain further information about the vector or axial-vector nature of massive gluon couplings. The differing collinear behavior of the cross sections allows to distinguish axigluon effects from QCD and other vector contributions. An axigluon generates a second maximum of $A_C^{|y|}$ in the forward-jet region, in addition to the common maximum for central jets. In order to exploit this information, however, a dedicated study of the collinear enhancement and a good experimental control of the measurement in the forward region are necessary.

At the full power of the LHC, $\sqrt{s} = 14\,\text{TeV}$, the integrated charge asymmetry is smaller than at $\sqrt{s} = 8\,\text{TeV}$ due to the increased $gg$ background. However, since the cross section for $t\bar t + j$ production is larger, stronger cuts can be applied. We found that the resulting charge asymmetry $A_C^{|y|}$ in QCD can be of the same size or larger than at LHC8 for comparable cross sections. For massive gluon searches, LHC14 can be a convenient setup because the large CM energy allows to produce very heavy such particles. LHC14 is therefore not disfavored to explore a charge asymmetry both in QCD and from massive gluons, and might be even more successful than LHC8 in the search for heavy resonances.

We conclude that the charge asymmetry in $t\bar t + j$ production provides an interesting playground for tests of QCD and of physics beyond the SM in top-quark pair production. The jet angular distribution serves as a useful tool to explore the kinematic features of the final state on top of those in inclusive $t\bar t$ production. The smallness of the QCD charge asymmetry in $t\bar t + j$ production at the LHC is a challenge for future measurements. In turn, it makes $A_C^{|y|}$ a clean observable to test the properties of new particles via charge conjugation. The year of the top quark is in its full bloom. Top-quark physics beyond inclusive $t\bar t$ production at the LHC is in its infancy, but will soon be testable in more detail.

\section{Acknowledgements}
It is our pleasure to thank Joe Boudreau, Andreas von Manteuffel, Lucia Masetti, Ben Pecjak and Hubert Spiesberger for helpful discussions and Stefan Weinzierl for comments on the manuscript. The work of S.~B. is supported by the Initiative and Networking Fund
of the Helmholtz Association, contract HA-101 (`Physics at the Terascale') and by the Research Center `Elementary Forces and Mathematical Foundations' of the Johannes Gutenberg-Universit\"at Mainz.

\begin{appendix}

\section{Phase space kinematics for $t\bar t + j$ production}\label{app:phasespace}
We parameterize the momenta of the initial-state hadrons by $P_1 = \sqrt{s}/2\left(x_1,0,0,x_1\right)$ and $P_2 = \sqrt{s}/2\left(x_2,0,0,-x_2\right)$. The boost $\beta$ of the partonic CM frame with respect to the lab frame can then be expressed in terms of the parton momentum fractions $x_1$ and $x_2$ as \footnote{Here, all momenta and energies are defined in the lab frame. The $z$ axis coincides with the direction of the incoming hadron with momentum $P_1$.}
 \begin{eqnarray}
\beta & = & \frac{p_{t}^{z}+p_{\bar t}^{z}+p_{j}^{z}}{E_{t}+E_{\bar t}+E_{j}} = \frac{P_1^z + P_2^z}{P_1^0 + P_2^0} = \frac{x_1-x_2}{x_1+x_2}\,,\qquad -1 < \beta < 1\,.
 \end{eqnarray}
Experimentally, the boost of the partonic CM frame can be determined by measuring the rapidity of the $t\bar t + j$ system in the lab frame, $y_{t\bar tj}$.\\[-25pt]
\begin{multicols}{2}
Both quantities are related via\\[0pt]
\begin{eqnarray}\label{eq:yttbj}
y_{t\bar t j} & = & \frac{1}{2}\ln\frac{\sum_{i=t,\bar t,j}(E_i + p_i^{z})}{\sum_{i=t,\bar t,j}(E_i - p_i^{z})}\\\nonumber
& \stackrel{\text{LO}}{=} & \frac{1}{2}\ln\frac{x_1}{x_2} = \frac{1}{2}\ln\frac{1+\beta}{1-\beta}\,.\\\nonumber
 & & \\\nonumber
\end{eqnarray}
%%%%%%%%%%%%%%%%%%%%%%%%%%%%%%%%%%%%%%%%%%%%%%%%%%%%%%%%%%%%%%%%
\begin{figurehere}
\hspace*{2cm}
\includegraphics[width=0.7\columnwidth]{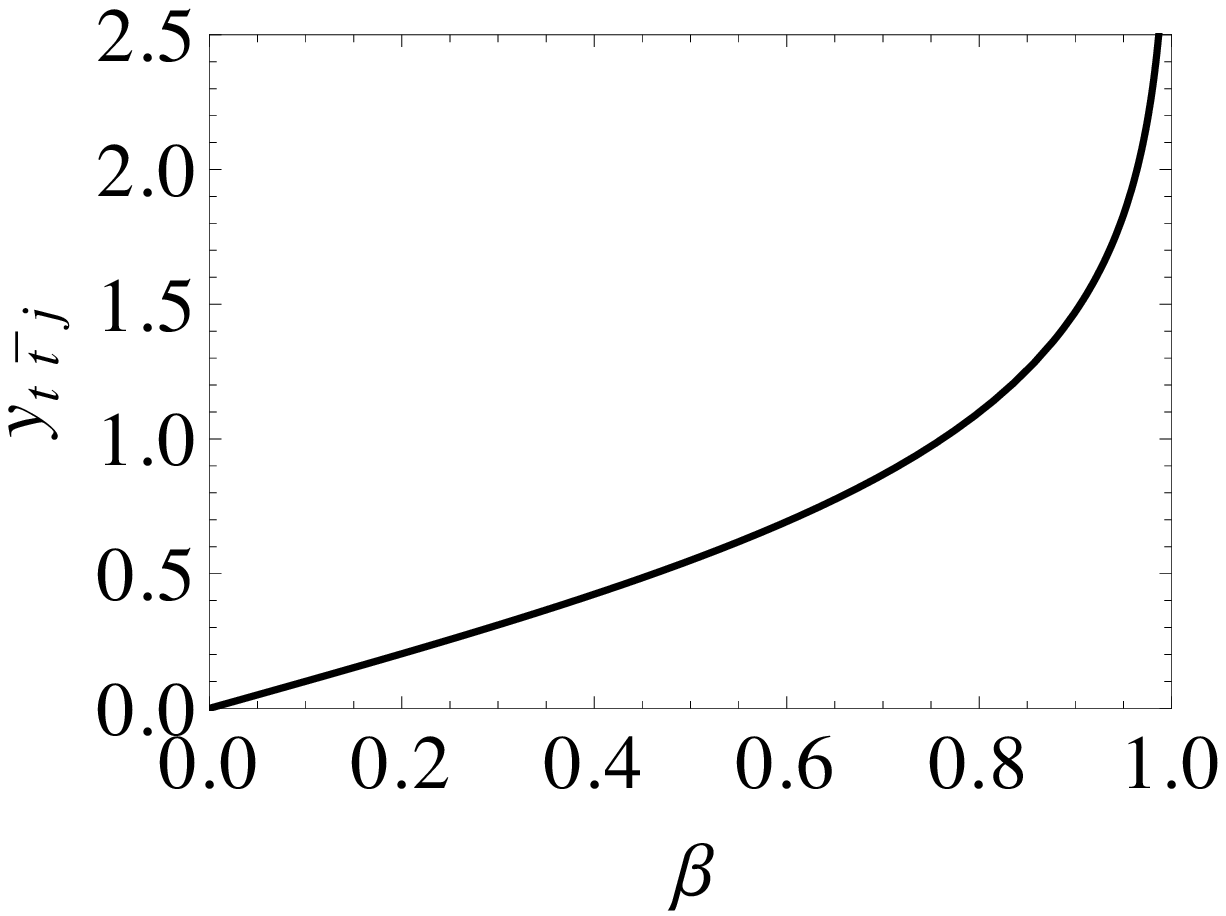}
\end{figurehere}
%%%%%%%%%%%%%%%%%%%%%%%%%%%%%%%%%%%%%%%%%%%%%%%%%%%%%%%%%%%%%%%%
\end{multicols}
\vspace*{-0.7cm}
A lower cut on the rapidity $y_{t\bar tj}$ in the lab frame therefore directly translates into a selection of events with a higher boost $\beta$.

\end{appendix}


\begin{thebibliography}{999}

%%%%%%%%%%%%%%%%%%%%%%%%%%%%%%%%%%%%%%%%%%%%%%%%%%%%%%
%ttbar cross section
%%%%%%%%%%%%%%%%%%%%%%%%%%%%%%%%%%%%%%%%%%%%%%%%%%%%%%

\bibitem{ATLAS:xsec} 
  G.~Aad {\it et al.}  [ATLAS Collaboration],
  %``Statistical combination of top quark pair production cross-section measurements using dilepton, single-lepton, and all-hadronic final states at sqrt{s} = 7 TeV with the ATLAS detector'',
  ATLAS-CONF-2012-024.

\bibitem{CMS:xsec} 
  S.~Chatrchyan {\it et al.}  [CMS Collaboration],
  %``Measurement of the t¯t production cross section in the dilepton channel in pp collisions at sqrt{s} = 8 TeV'',
  CMS-PAS-TOP-12-007.

%%%%%%%%%%%%%%%%%%%%%%%%%%%%%%%%%%%%%%%%%%%%%%%%%%%%%%

\bibitem{Langenfeld:2009wd} 
  U.~Langenfeld, S.~Moch and P.~Uwer,
  %``Measuring the running top-quark mass,''
  Phys.\ Rev.\ D {\bf 80}, 054009 (2009)
  [arXiv:0906.5273 [hep-ph]].

\bibitem{Ahrens:2011px} 
  V.~Ahrens, A.~Ferroglia, M.~Neubert, B.~D.~Pecjak and L.~L.~Yang,
  %``Precision predictions for the t+t(bar) production cross section at hadron colliders,''
  Phys.\ Lett.\ B {\bf 703}, 135 (2011)
  [arXiv:1105.5824 [hep-ph]].

%%%%%%%%%%%%%%%%%%%%%%%%%%%%%%%%%%%%%%%%%%%%%%%%%%%%%%
%ttbar resonance searches
%%%%%%%%%%%%%%%%%%%%%%%%%%%%%%%%%%%%%%%%%%%%%%%%%%%%%%
\bibitem{Aaltonen:2011ts} 
  T.~Aaltonen {\it et al.}  [CDF Collaboration],
  %``A Search for resonant production of $t\bar{t}$ pairs in $4.8\ \rm{fb}^{-1}$ of integrated luminosity of $p\bar{p}$ collisions at $\sqrt{s}=1.96\ \rm{TeV}$,''
  Phys.\ Rev.\ D {\bf 84}, 072004 (2011)
  [arXiv:1107.5063 [hep-ex]].

\bibitem{Abazov:2011gv} 
  V.~M.~Abazov {\it et al.}  [D0 Collaboration],
  %``Search for a Narrow $t\bar{t}$ Resonance in $p\bar{p}$ Collisions at $\sqrt{s}=1.96$ TeV,''
  Phys.\ Rev.\ D {\bf 85}, 051101 (2012)
  [arXiv:1111.1271 [hep-ex]].

\bibitem{ATLAS:2012tx} 
  G.~Aad {\it et al.}  [ATLAS Collaboration],
  %``A search for $t\bar{t}$ resonances in lepton+jets events with highly boosted top quarks collected in $pp$ collisions at $\sqrt{s} = 7$ TeV with the ATLAS detector,''
  JHEP {\bf 1209}, 041 (2012)
  [arXiv:1207.2409 [hep-ex]].

\bibitem{CMS:ttboost} 
  S.~Chatrchyan {\it et al.}  [CMS Collaboration],
  %``Search for anomalous t t-bar production in the highly-boosted all-hadronic final state,''
  arXiv:1204.2488 [hep-ex].
%%%%%%%%%%%%%%%%%%%%%%%%%%%%%%%%%%%%%%%%%%%%%%%%%%%%%%

%%%%%%%%%%%%%%%%%%%%%%%%%%%%%%%%%%%%%%%%%%%%%%%%%%%%%%
%inclusive top charge asymmetry
%%%%%%%%%%%%%%%%%%%%%%%%%%%%%%%%%%%%%%%%%%%%%%%%%%%%%%
\bibitem{CDF:afbt}
  T.~Aaltonen {\it et al.}  [CDF~Collaboration], CDF Public Note 10807 (2012);
  %``Evidence for a Mass Dependent Forward-Backward Asymmetry in Top Quark Pair Production,''
  Phys.\ Rev.\ D {\bf 83}, 112003 (2011)
  [arXiv:1101.0034 [hep-ex]]; CDF Public Note 10436 (2011).

\bibitem{Abazov:2011rq} 
  V.~M.~Abazov {\it et al.}  [D0 Collaboration],
  %``Forward-backward asymmetry in top quark-antiquark production,''
  Phys.\ Rev.\ D {\bf 84}, 112005 (2011)
  [arXiv:1107.4995 [hep-ex]].

\bibitem{Kuhn:2011ri} 
  J.~H.~K\"uhn and G.~Rodrigo,
  %``Charge asymmetries of top quarks at hadron colliders revisited,''
  JHEP {\bf 1201}, 063 (2012)
  [arXiv:1109.6830 [hep-ph]].

\bibitem{Ahrens:2011uf} 
  V.~Ahrens, A.~Ferroglia, M.~Neubert, B.~D.~Pecjak and L.~L.~Yang,
  %``The top-pair forward-backward asymmetry beyond NLO,''
  Phys.\ Rev.\ D {\bf 84}, 074004 (2011)
  [arXiv:1106.6051 [hep-ph]].

\bibitem{ATLAS:ac}
  G.~Aad {\it et al.}  [ATLAS Collaboration],
  %``Measurement of the charge asymmetry in dileptonic decays of top quark pairs in pp collisions at √s = 7 TeV using the ATLAS detector'',
  ATLAS-CONF-2012-057.

\bibitem{CMS:ac} 
  S.~Chatrchyan {\it et al.}  [CMS Collaboration],
  %``Differential measurements of the charge asymmetry in top quark pair production'',
  CMS-PAS-TOP-11-030.
%%%%%%%%%%%%%%%%%%%%%%%%%%%%%%%%%%%%%%%%%%%%%%%%%%%%%%

\bibitem{Rainwater:1999sd} 
  D.~L.~Rainwater and D.~Zeppenfeld,
  %``Observing H ---> W(*) W(*) ---> e+- muon-+ missing-p(T) in weak boson fusion with dual forward jet tagging at the CERN LHC,''
  Phys.\ Rev.\ D {\bf 60}, 113004 (1999)
  [Erratum-ibid.\ D {\bf 61}, 099901 (2000)]
  [hep-ph/9906218].

\bibitem{Mangano:2008ha} 
  M.~L.~Mangano,
  %``Standard Model backgrounds to supersymmetry searches,''
  Eur.\ Phys.\ J.\ C {\bf 59}, 373 (2009)
  [arXiv:0809.1567 [hep-ph]].

\bibitem{Dittmaier:2007wz} 
  S.~Dittmaier, P.~Uwer and S.~Weinzierl,
  %``NLO QCD corrections to t anti-t + jet production at hadron colliders,''
  Phys.\ Rev.\ Lett.\  {\bf 98}, 262002 (2007)
  [hep-ph/0703120 [hep-ph]].

\bibitem{Dittmaier:2008uj} 
  S.~Dittmaier, P.~Uwer and S.~Weinzierl,
  %``Hadronic top-quark pair production in association with a hard jet at next-to-leading order QCD: Phenomenological studies for the Tevatron and the LHC,''
  Eur.\ Phys.\ J.\ C {\bf 59}, 625 (2009)
  [arXiv:0810.0452 [hep-ph]].

\bibitem{Melnikov:2010iu} 
  K.~Melnikov and M.~Schulze,
  %``NLO QCD corrections to top quark pair production in association with one hard jet at hadron colliders,''
  Nucl.\ Phys.\ B {\bf 840}, 129 (2010)
  [arXiv:1004.3284 [hep-ph]].

\bibitem{ATLAS:2012ttbj} 
  G.~Aad {\it et al.}  [ATLAS Collaboration],
  %``Measurement of the cross section for tt̄+jets production using a kinematic fit method with the ATLAS detector,''
  ATLAS-CONF-2012-083.

\bibitem{Melnikov:2011qx} 
  K.~Melnikov, A.~Scharf and M.~Schulze,
  %``Top quark pair production in association with a jet: QCD corrections and jet radiation in top quark decays,''
  Phys.\ Rev.\ D {\bf 85}, 054002 (2012)
  [arXiv:1111.4991 [hep-ph]].

\bibitem{Kardos:2011qa} 
  A.~Kardos, C.~Papadopoulos and Z.~Trocsanyi,
  %``Top quark pair production in association with a jet with NLO parton showering,''
  Phys.\ Lett.\ B {\bf 705}, 76 (2011)
  [arXiv:1101.2672 [hep-ph]].

\bibitem{Alioli:2011as} 
  S.~Alioli, S.~Moch and P.~Uwer,
  %``Hadronic top-quark pair-production with one jet and parton showering,''
  JHEP {\bf 1201}, 137 (2012)
  [arXiv:1110.5251 [hep-ph]].

\bibitem{Chatrchyan:2012su} 
  S.~Chatrchyan {\it et al.}  [CMS Collaboration],
  %``Search for charge-asymmetric production of W' bosons in top pair + jet events from pp collisions at sqrt(s) = 7 TeV,''
  arXiv:1206.3921 [hep-ex].

\bibitem{Ferrario:2009ee} 
  P.~Ferrario and G.~Rodrigo,
  %``Heavy colored resonances in t t-bar + jet at the LHC,''
  JHEP {\bf 1002}, 051 (2010)
  [arXiv:0912.0687 [hep-ph]].

\bibitem{Halzen:1987xd} 
  F.~Halzen, P.~Hoyer and C.~S.~Kim,
  %``Forward - Backward Asymmetry Of Hadroproduced Heavy Quarks In Qcd,''
  Phys.\ Lett.\ B {\bf 195}, 74 (1987).

\bibitem{AguilarSaavedra:2011cp} 
  J.~A.~Aguilar-Saavedra, A.~Juste and F.~Rubbo,
  %``Boosting the t tbar charge asymmetry,''
  Phys.\ Lett.\ B {\bf 707}, 92 (2012)
  [arXiv:1109.3710 [hep-ph]].

\bibitem{Lancaster:2011wr} 
  [Tevatron Electroweak Working Group and CDF and D0 Collaborations],
  %``Combination of CDF and D0 results on the mass of the top quark using up to 5.8~fb-1 of data,''
  arXiv:1107.5255 [hep-ex].

\bibitem{Pumplin:2002vw} 
  J.~Pumplin, D.~R.~Stump, J.~Huston, H.~L.~Lai, P.~M.~Nadolsky and W.~K.~Tung,
  %``New generation of parton distributions with uncertainties from global QCD analysis,''
  JHEP {\bf 0207}, 012 (2002) [hep-ph/0201195].

\bibitem{Hahn:2004fe} 
  T.~Hahn,
  %``CUBA: A Library for multidimensional numerical integration,''
  Comput.\ Phys.\ Commun.\  {\bf 168}, 78 (2005) [hep-ph/0404043].

\bibitem{Galassi} 
  M. Galassi {\it et al.}, GNU Scientific Library Reference Manual (3rd ed.), ISBN 0954612078,
  \hyperref[http://www.gnu.org/software/gsl/]{http://www.gnu.org/software/gsl/}.

\bibitem{CDF:2009}
  T.~Aaltonen {\it et al.}  [CDF~Collaboration], CDF Public Note 9850 (2009).

\bibitem{Frampton:1987dn} 
  P.~H.~Frampton and S.~L.~Glashow,
  %``Chiral Color: An Alternative to the Standard Model,''
  Phys.\ Lett.\ B {\bf 190}, 157 (1987).

\bibitem{Hill:1991at} 
  C.~T.~Hill,
  %``Topcolor: Top quark condensation in a gauge extension of the standard model,''
  Phys.\ Lett.\ B {\bf 266}, 419 (1991).

\bibitem{Chivukula:2011ng} 
  R.~S.~Chivukula, A.~Farzinnia, E.~H.~Simmons and R.~Foadi,
  %``Production of Massive Color-Octet Vector Bosons at Next-to-Leading Order,''
  Phys.\ Rev.\ D {\bf 85}, 054005 (2012)
  [arXiv:1111.7261 [hep-ph]].

\bibitem{Lillie:2007yh} 
  B.~Lillie, L.~Randall and L.~-T.~Wang,
  %``The Bulk RS KK-gluon at the LHC,''
  JHEP {\bf 0709}, 074 (2007) [hep-ph/0701166].

\bibitem{Kuhn:1998kw} 
  J.~H.~K\"uhn and G.~Rodrigo,
  %``Charge asymmetry of heavy quarks at hadron colliders,''
  Phys.\ Rev.\ D {\bf 59}, 054017 (1999) [hep-ph/9807420].

\bibitem{ATLAS:dijets}
  G.~Aad {\it et al.} [ATLAS Collaboration], ATLAS-CONF-2012-088.

\bibitem{CMS:dijets}
  S.~Chatrchyan {\it et al.} [CMS Collaboration], CMS-PAS-EXO-12-016.

\bibitem{Haisch:2011up} 
  U.~Haisch and S.~Westhoff,
  %``Massive Color-Octet Bosons: Bounds on Effects in Top-Quark Pair Production,''
  JHEP {\bf 1108}, 088 (2011) [arXiv:1106.0529 [hep-ph]].\\
  Update: S.~Westhoff, talk given at Planck 2012, Warsaw.

\bibitem{Krohn:2011tw} 
  D.~Krohn, T.~Liu, J.~Shelton and L.~-T.~Wang,
  %``A Polarized View of the Top Asymmetry,''
  Phys.\ Rev.\ D {\bf 84}, 074034 (2011)
  [arXiv:1105.3743 [hep-ph]].

\bibitem{ATLAS:toppol}
  G.~Aad {\it et al.} [ATLAS Collaboration], ATLAS-CONF-2012-133.

\bibitem{CMS:toppol}
  S.~Chatrchyan {\it et al.} [CMS Collaboration], CMS-PAS-TOP-12-016.

\bibitem{ATLAS:2012ao} 
  G.~Aad {\it et al.}  [ATLAS Collaboration],
  %``Observation of spin correlation in $t \bar{t}$ events from pp collisions at sqrt(s) = 7 TeV using the ATLAS detector,''
  Phys.\ Rev.\ Lett.\  {\bf 108}, 212001 (2012)
  [arXiv:1203.4081 [hep-ex]].

\bibitem{CMS:spincorr}
  S.~Chatrchyan {\it et al.} [CMS Collaboration], CMS-PAS-TOP-12-004.

\bibitem{Frampton:2009rk} 
  P.~H.~Frampton, J.~Shu and K.~Wang,
  %``Axigluon as Possible Explanation for p anti-p ---> t anti-t Forward-Backward Asymmetry,''
  Phys.\ Lett.\ B {\bf 683}, 294 (2010)
  [arXiv:0911.2955 [hep-ph]].

\bibitem{Aaltonen:2009iz} 
  T.~Aaltonen {\it et al.}  [CDF Collaboration],
  %``First Measurement of the t anti-t Differential Cross Section d sigma/dM(t anti-t) in p anti-p Collisions at s**(1/2)=1.96-TeV,''
  Phys.\ Rev.\ Lett.\  {\bf 102}, 222003 (2009)
  [arXiv:0903.2850 [hep-ex]].

\bibitem{ATLAS:2012hg} 
  G.~Aad {\it et al.}  [ATLAS Collaboration],
  %``Measurements of top quark pair relative differential cross-sections with ATLAS in pp collisions at sqrt(s) = 7 TeV,''
  arXiv:1207.5644 [hep-ex].

\bibitem{CMS:ttbspectrum}
  S.~Chatrchyan {\it et al.} [CMS Collaboration], CMS-PAS-TOP-12-013.

\bibitem{Tavares:2011zg} 
  G.~M.~Tavares and M.~Schmaltz,
  %``Explaining the t-tbar asymmetry with a light axigluon,''
  Phys.\ Rev.\ D {\bf 84}, 054008 (2011)
  [arXiv:1107.0978 [hep-ph]].

%%%%%%%%%%%%%%%%%%%%%%%%%%%%%%%%%%%%%%%%%%%%%%%%%%%%%%%%%%%%%%%%%%%%%%%%%%%%%%%%%%%%%

\end{thebibliography}
\end{document}